\documentclass[aps,prd,final,reprint,showkeys,showpacs,superscriptaddress,floatfix,nofootinbib,twocolumn,amsmath,amssymb]{revtex4-1}

\usepackage[utf8]{inputenc}

\usepackage{hyperref} 
\usepackage{url}
\usepackage{graphicx}

\usepackage{graphics}
\usepackage{natbib}
\graphicspath{{./}}

\usepackage[normalem]{ulem}
\usepackage{color}

\usepackage{soul}

\def\units#1{~\hbox{$\,{\rm #1}$}}

\begin{document}

\title{Cosmic-ray interactions with the Sun using the {\tt FLUKA} code}
\author{M.~N.~Mazziotta}
\email[Corresponding author ]{mazziotta@ba.infn.it}
\homepage{https://orcid.org/0000-0001-9325-4672}
\affiliation{Istituto Nazionale di Fisica Nucleare, Sezione di Bari, via Orabona 4, I-70126 Bari, Italy}
\author{P.~De~La~Torre~Luque}
\affiliation{Istituto Nazionale di Fisica Nucleare, Sezione di Bari, via Orabona 4, I-70126 Bari, Italy}
\affiliation{Dipartimento di Fisica ``M. Merlin" dell'Universit\`a e del Politecnico di Bari, via Amendola 173, I-70126 Bari, Italy}
\author{L.~Di~Venere}
\affiliation{Istituto Nazionale di Fisica Nucleare, Sezione di Bari, I-70126 Bari, Italy}
\author{A.~Fass\`o}
\affiliation{13 Passage Hamo, CH-1262 EYSINS, Switzerland}
\author{A.~Ferrari}
\affiliation{CERN, the European Organization for Nuclear Research, Esplanade des Particules 1, 1211 Geneva, Switzerland}
\author{F.~Loparco}
\homepage{https://orcid.org/0000-0002-1173-5673}
\affiliation{Istituto Nazionale di Fisica Nucleare, Sezione di Bari, via Orabona 4, I-70126 Bari, Italy}
\affiliation{Dipartimento di Fisica ``M. Merlin" dell'Universit\`a e del Politecnico di Bari, via Amendola 173, I-70126 Bari, Italy}
\author{P.~R.~Sala}
\affiliation{Istituto Nazionale di Fisica Nucleare, Sezione di Milano, Via Celoria,16, 20133 Milano, Italy}
\author{D.~Serini}
\homepage{https://orcid.org/0000-0002-9754-6530}
\affiliation{Istituto Nazionale di Fisica Nucleare, Sezione di Bari, via Orabona 4, I-70126 Bari, Italy}
\affiliation{Dipartimento di Fisica ``M. Merlin" dell'Universit\`a e del Politecnico di Bari, via Amendola 173, I-70126 Bari, Italy}

\date{\today}

\begin{abstract}
The interactions of cosmic rays with the solar atmosphere produce secondary particle which can reach the Earth. In this work we present a comprehensive calculation of the yields of secondary particles as gamma-rays, electrons, positrons, neutrons and neutrinos performed
with the {\tt FLUKA} code. We also estimate the intensity at the Sun and the fluxes at the Earth of these secondary particles by folding their yields with the intensities of cosmic rays impinging on the solar surface. The results are sensitive on the assumptions on the magnetic field nearby the Sun and to the cosmic-ray transport in the magnetic field in the inner solar system. 
\end{abstract}

\maketitle

\section{Introduction}

Cosmic rays entering in the Solar system after propagating for millions of years in the Galaxy can reach the planets and the Sun itself, producing emission of secondary particles, such as gamma rays and neutrinos, due to the interactions with the surfaces or the atmospheres of the celestial bodies. 

The Moon~\cite{Abdo:2012nfa,Cerutti:2016gts} and the Earth~\cite{Abdo:2009gt} are both bright sources of gamma rays. Lunar and terrestrial gamma rays are originated from the hadronic interactions of cosmic-ray nuclei with the lunar surface and with the upper layers of the Earth's atmosphere, respectively. The Sun is also a bright source of high-energy gamma rays. While gamma rays from the Earth and from the Moon are originated from cosmic-ray nuclei, the solar gamma-ray emission consists of two components: the first one, called \textit{disk emission}, is due to cosmic-ray nuclei interacting with the solar surface~\cite{Abdo:2011xn,Seckel:1991ffa}, and is localized around the solar disk; the second one, which is due to the inverse Compton scatterings of cosmic-ray electrons (and positrons) with the solar optical photons, is a diffuse component and extends up to tens of degrees from the Sun~\cite{Abdo:2011xn,Orlando:2008uk,Orlando:2006zs,Moskalenko:2006ta}.

Several attempts have already been done to calculate the secondary emission (e.g., gamma rays and neutrinos) due to the interactions of cosmic rays with the solar atmosphere (see for example~\cite{Seckel:1991ffa,Edsjo:2017kjk}). In particular, the knowledge of such emission could be used to constrain exotic processes, such as the production of standard model particles in the annihilation of dark matter particles captured by the Sun~\cite{Edsjo:2017kjk,Ng:2017aur,Zhou:2016ljf,Masip:2017gvw,Cuoco:2019mlb}.

Early predictions are based on semi-analytical calculations with the inclusion of solar magnetic field~\cite{Seckel:1991ffa}, while full numerical simulations for the production of neutrinos based on the Monte Carlo method have been performed in \cite{Ingelman:1996mj} and recently revisited and updated by Refs.~\cite{Edsjo:2017kjk}. However, in those Monte Carlo simulations, the effect of the magnetic field was neglected since the calculation was performed at high energies. The production of neutrinos is closely related to that of gamma rays in the solar disk, as both are originated from hadronic interactions of cosmic-ray nuclei.

In this work we have performed a full simulation with the {\tt FLUKA} code to calculate the yields of secondary particles produced by the interactions of cosmic rays with the Sun. In particular, we have simulated the interactions of protons, helium nuclei and electrons impinging on the solar atmosphere in a wide energy range from 0.1\units{GeV/n} to 100\units{TeV/n}, while the energy of secondary particles has been simulated down to 100\units{keV}. The low-energy region is extremely interesting for the proposed future gamma-ray telescopes~\cite{DeAngelis:2016slk,DeAngelis:2017gra,mcenery2019allsky}, which aim to probe photon energy intervals extending well below the lower bound of that explored by the Fermi Large Area Telescope (a few tens of $\units{MeV}$)~\cite{Atwood:2009ez}.

The present work is based on our previous ones, in which we evaluated, using {\tt FLUKA}, the yields of secondary cosmic rays in the collisions of primary cosmic rays with the interstellar gas~\cite{Mazziotta:2015uba} and the lunar gamma-ray emission~\cite{Cerutti:2016gts}.

In simulating the interactions of cosmic rays with the solar atmosphere there are a number of important effects to be considered. First, the interplanetary magnetic field affects the spectra of cosmic rays reaching the Sun. Second, the strong heliospheric magnetic field nearby the Sun also affects the trajectories of charged particles in the solar atmosphere: in particular, the total path length increases with the intensity of the magnetic field, and this corresponds to an increase of the interaction probability and consequently to an increase of the cascades of secondary particles. Finally, the profile of the solar atmosphere needs to be accounted in detail, since the cascades usually develop from a low-density medium toward a denser medium; in addition, the yield of secondary particles far away from the Sun is also affected by the grammage along the line of the sight from the production point to the outer space.

\section{Simulation set-up}
\label{sec:simu}

The propagation and the interactions of cosmic rays with the solar atmosphere have been simulated with the {\tt FLUKA} code~\cite{Ferrari:2005zk,BOHLEN2014211,BATTISTONI201510}. {\tt FLUKA} is a general purpose Monte Carlo code for the simulation of hadronic and electromagnetic interactions, used in many applications. It can simulate with high accuracy the interactions and propagation in matter of about 60 different species of particles, including photons and electrons from $1\units{keV}$ to thousands of $\units{TeV}$, neutrinos, muons of any energy, hadrons of energies up to $20\units{TeV}$ (up to $10\units{PeV}$ when it is interfaced with the {\tt DPMJET} code~\cite{10.1007/978-3-642-18211-2_166}) and all the corresponding antiparticles, neutrons down to thermal energies and heavy ions.

Hadronic interactions are treated in {\tt FLUKA} following a theory-driven approach. Below a few$\units{GeV}$, the hadron-nucleon interaction model is based on resonance production and decay of particles, while for higher energies the Dual Parton Model (DPM) is used, implying a treatment in terms of quark chain formation and hadronization. The extension from hadron-nucleon to hadron-nucleus interactions is done in the framework of the PreEquilibrium Approach to NUclear Thermalization model ({\tt PEANUT})~\cite{Fasso:2000hd,battistoni2006recent}, including the Gribov-Glauber multi-collision mechanism followed by the pre-equilibrium stage and eventually equilibrium processes (evaporation, fission, Fermi break-up and gamma deexcitation). 

{\tt FLUKA} can handle even very complex geometries, using an improved version of the well known Combinatorial Geometry (CG) package, that has been designed to track correctly both neutral and charged particles, even in the presence of magnetic fields.

In our code we use a spherical reference frame centered on the Sun, which is described as a sphere of radius $R_\odot=6.9551 \times 10^{10}\units{cm}$. The polar axis (i.e. z-axis) of the reference frame corresponds to the Sun's rotation axis.  Our simulation includes the radial profiles of the  chemical composition, of the density, of the temperature and of the pressure of the Sun (see Sec.~\ref{comp}). In addition, we have implemented various models of the magnetic field in the region close to the Sun (inner magnetic field, see Sec.~\ref{sec:inmf}), while for the interplanetary magnetic field we have used the Parker model (see Sec.~\ref{sec:imf}). As will be discussed in the next sections, the inner magnetic field affects the cosmic-ray interactions with the solar environment, while the interplanetary magnetic field affects their propagation to the Sun.  

To evaluate the yields of secondary particles from the Sun we have simulated several samples of protons, electrons and $^{4}$He nuclei with different kinetic energies impinging a sphere of radius $R_{SS}=2.5R_{\odot}$ surrounding the Sun, with an isotropic and uniform distribution. As it will be shown in Secs.~\ref{sec:inmf} and~\ref{sec:imf}, the generation sphere corresponds to the boundary between the inner and outer magnetic field regions. 
The primary kinetic energy values are taken on a grid of $97$ equally spaced values in a logarithmic scale, from $100\units{MeV/n}$ up to $100\units{TeV/n}$. 

\begin{figure}[!ht]
    \centering
    \includegraphics[width=0.5\textwidth,height=0.25\textheight,clip]{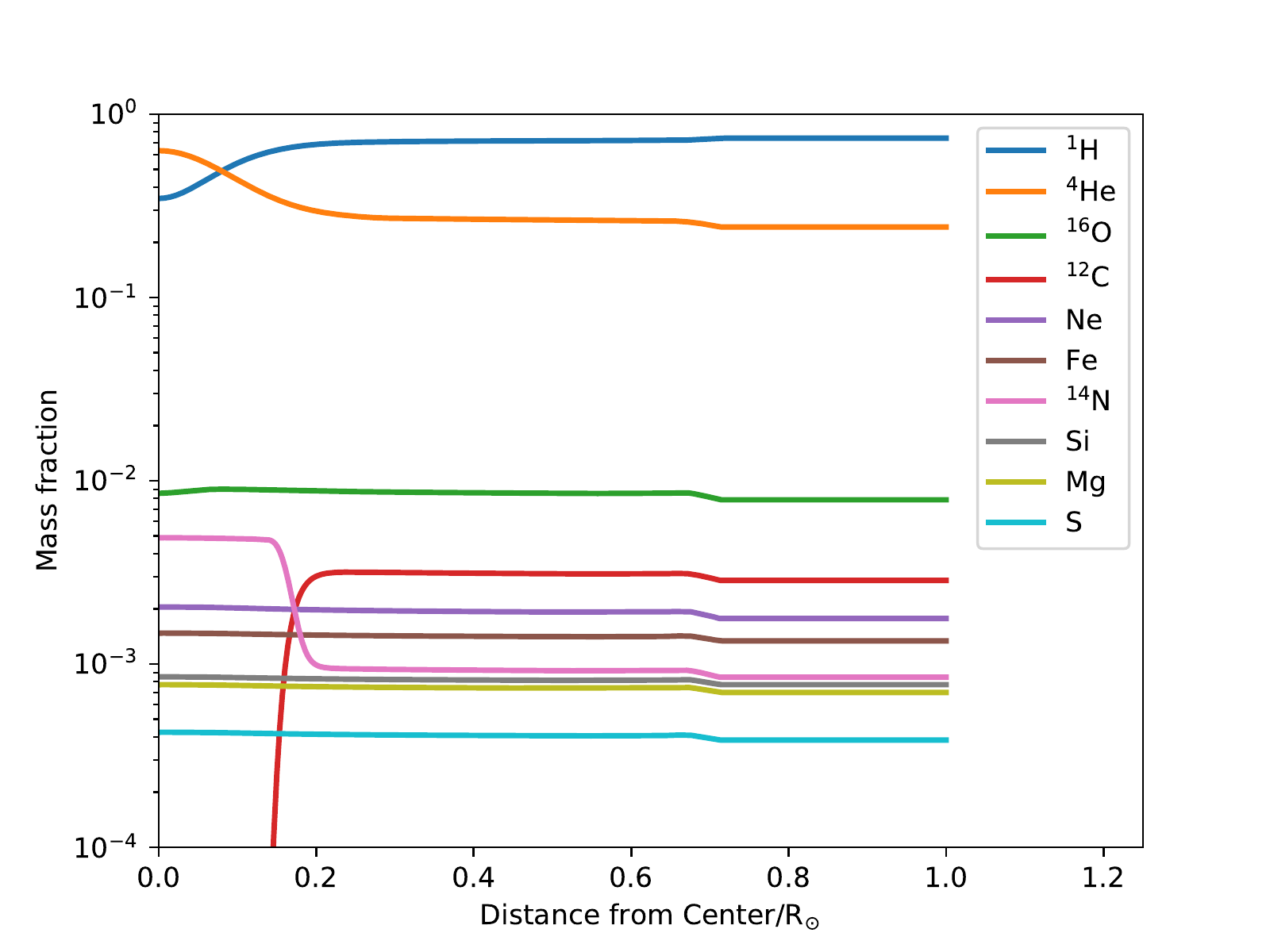}
    \caption{Mass fractions in the {\tt gs98} model as a function of the distance from the center of the Sun in units of $R_\odot$. Only atoms with mass fractions above $10^{-4}$ at the surface are shown. The data are taken from Ref.~\cite{Vinyoles:2016djt}.}
    \label{fig:compos}
\end{figure}

\subsection{Solar composition}
\label{comp}

In our simulation we have implemented a chemical composition profile of the Sun derived from the Standard Solar Models (SSMs) for the interior of the Sun, provided by Ref.~\cite{Vinyoles:2016djt} (hereafter Model {\tt gs98})\footnote{We use the data file \url{http://www.ice.csic.es/personal/aldos/Solar_Data_files/struct_b16_gs98.dat}}. Figure~\ref{fig:compos} shows the mass fractions of the main components as a function of the distance from the center of the Sun for the {\tt gs98} model. The main components are the hydrogen and $^{4}$He, while the abundances of heavier isotopes are below 1\%. Since most of cosmic-ray interactions will take place in the solar atmosphere, close to the surface of the Sun, we have extrapolated this model outside the Sun assuming that the chemical composition of the atmosphere is the same as that at $r=R_{\odot}$.

\begin{figure}[!ht]
\centering
\includegraphics[width=0.45\textwidth,height=0.2\textheight,clip]{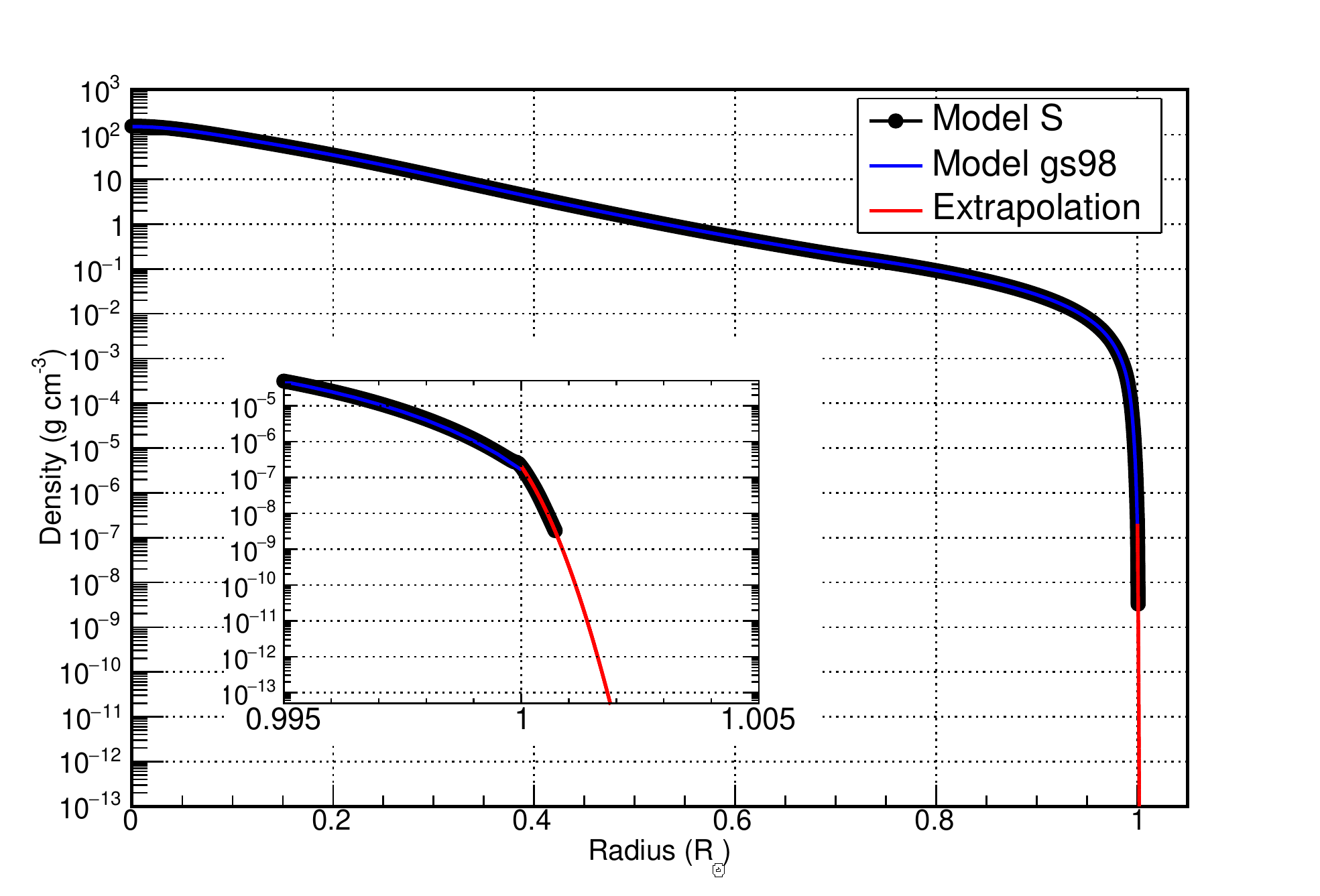}
\includegraphics[width=0.45\textwidth,height=0.2\textheight,clip]{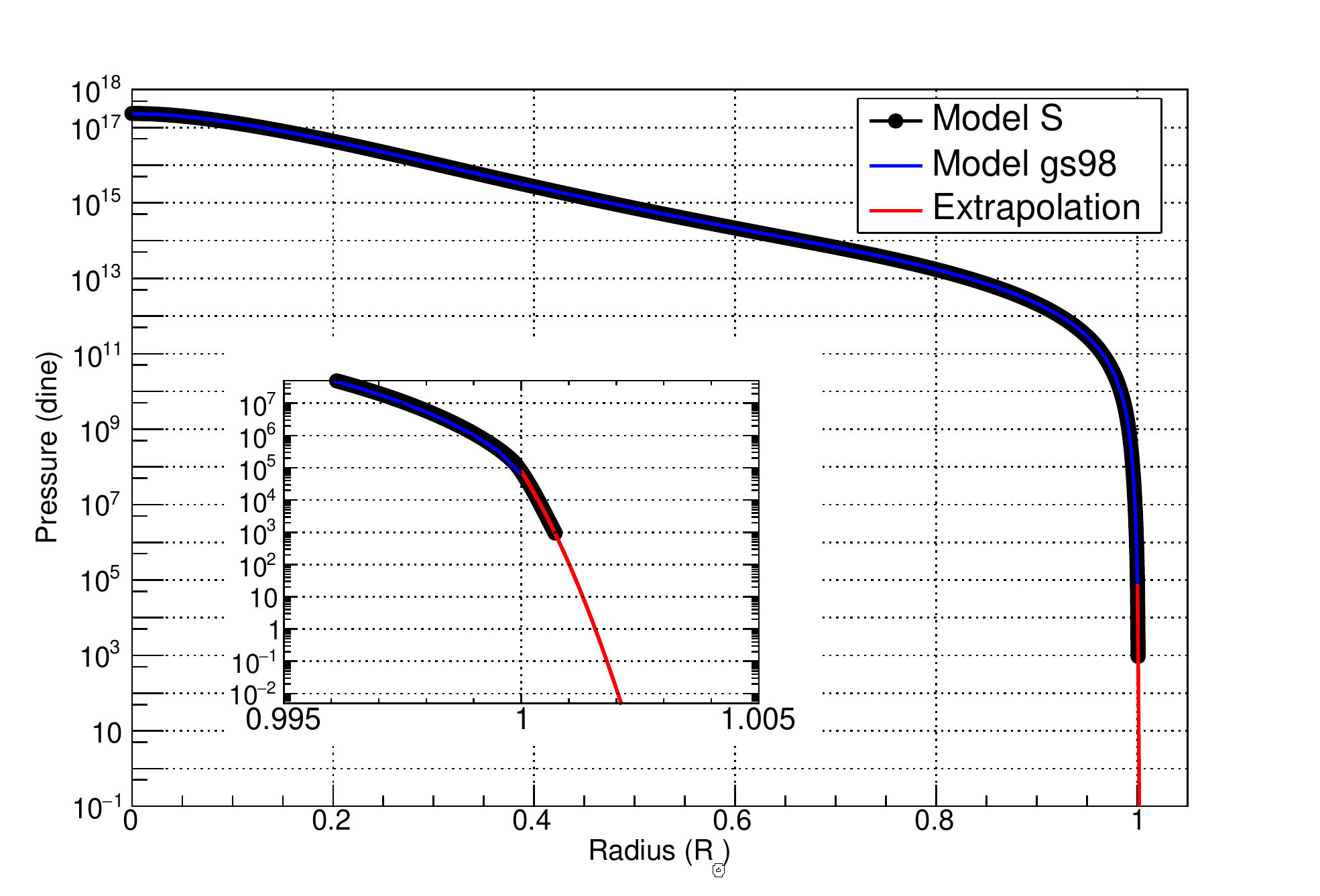} 
\includegraphics[width=0.45\textwidth,height=0.2\textheight,clip]{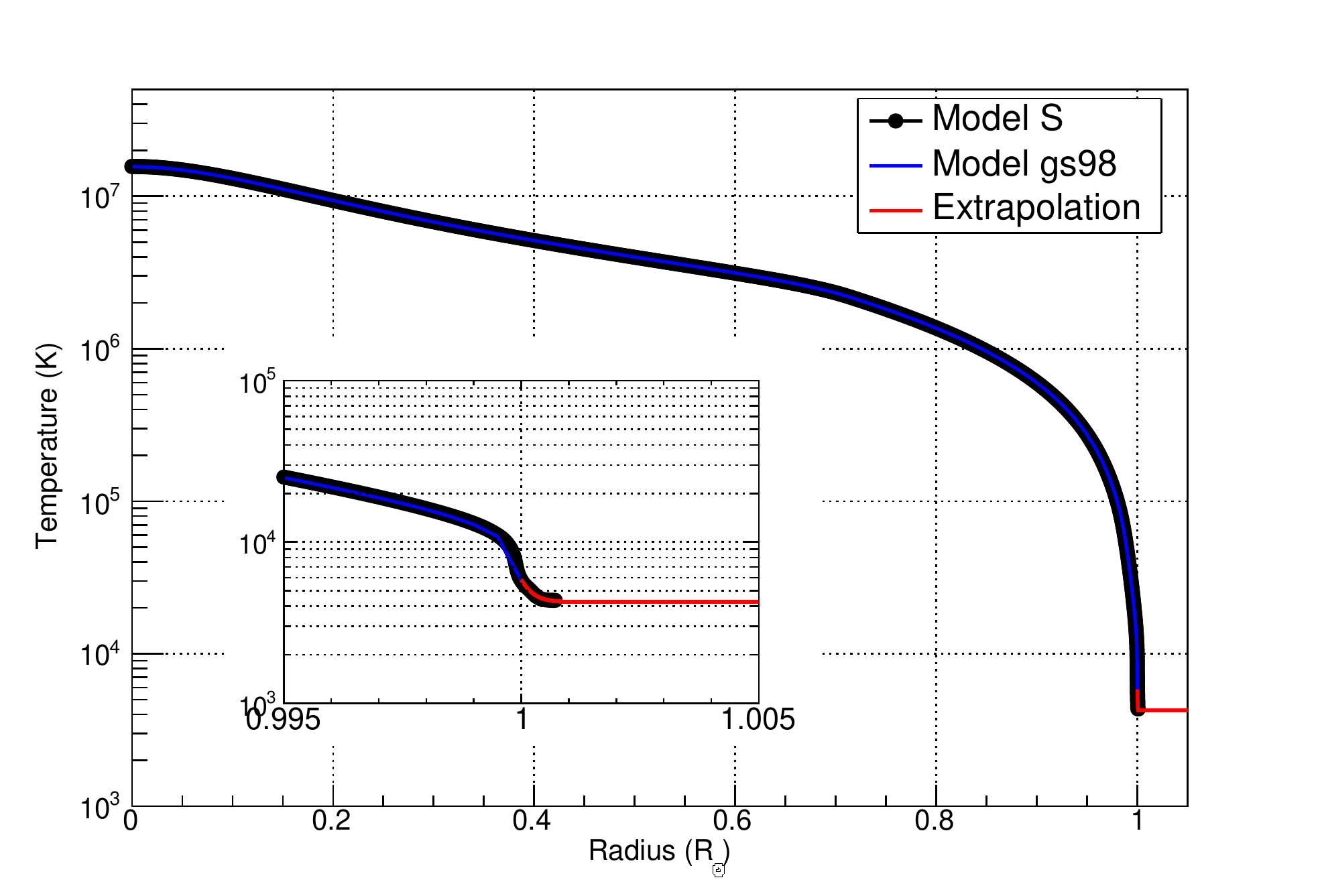} 
\caption{Density (top panel), pressure (middle panel) and temperature (bottom panel) as a function of the radial distance from the Sun centre in units of $R_\odot$. The black dots indicate the Model S~\cite{ChristensenDalsgaard:1996ap}; the blue line is the Model {\tt gs98}~\cite{Vinyoles:2016djt}; the red line is the present extrapolation. The inset shows a zoom near the Sun radius.}
\label{fig:profile}
\end{figure}

For the radial profiles of density, temperature and pressure we use the model provided by Ref.~\cite{ChristensenDalsgaard:1996ap} (hereafter Model S), since it extends up to about $500\units{km}$ above the solar surface. We then extrapolate this model to higher altitudes, up to about $1400\units{km}$. We have also verified that the Model {\tt gs98} is very similar to the Model S up to $r=R_{\odot}$.

Figure~\ref{fig:profile} shows the radial density (top panel), pressure (middle panel) and temperature (bottom panel) profiles. The Model S is shown with black points, the Model {\tt gs98} is shown with blue lines, and the extrapolation is shown with red lines.

In the {\tt FLUKA} simulation set-up we have implemented 100 layers (i.e. shells) with different densities and chemical compositions, divided in three sets equally spaced on a logarithmic density scale: the external 40 layers from about $10^{-13}\units{g/cm^3}$ up to $10^{-3}\units{g/cm^3}$, the middle 40 layers from $10^{-3}\units{g/cm^3}$ to $10^{-1}\units{g/cm^3}$ and the inner 20 layers with density $> 10^{-1}\units{g/cm^3}$. In each shell we define a compound mixture material according to the mass composition, density, temperature and pressure profiles shown in figures~\ref{fig:compos} and~\ref{fig:profile}. We have also implemented the temperature profile, since the neutron cross sections for the main isotopes (i.e. H, $^3$He, $^4$He and $^{12}$C) are dependent on the temperature. In particular, the temperature has an effect in the capture of neutrons that produce the gamma-ray line of $2.2\units{MeV}$. 

\begin{figure}[!ht]
\centering
\includegraphics[width=0.99\columnwidth,height=0.23\textheight,clip]{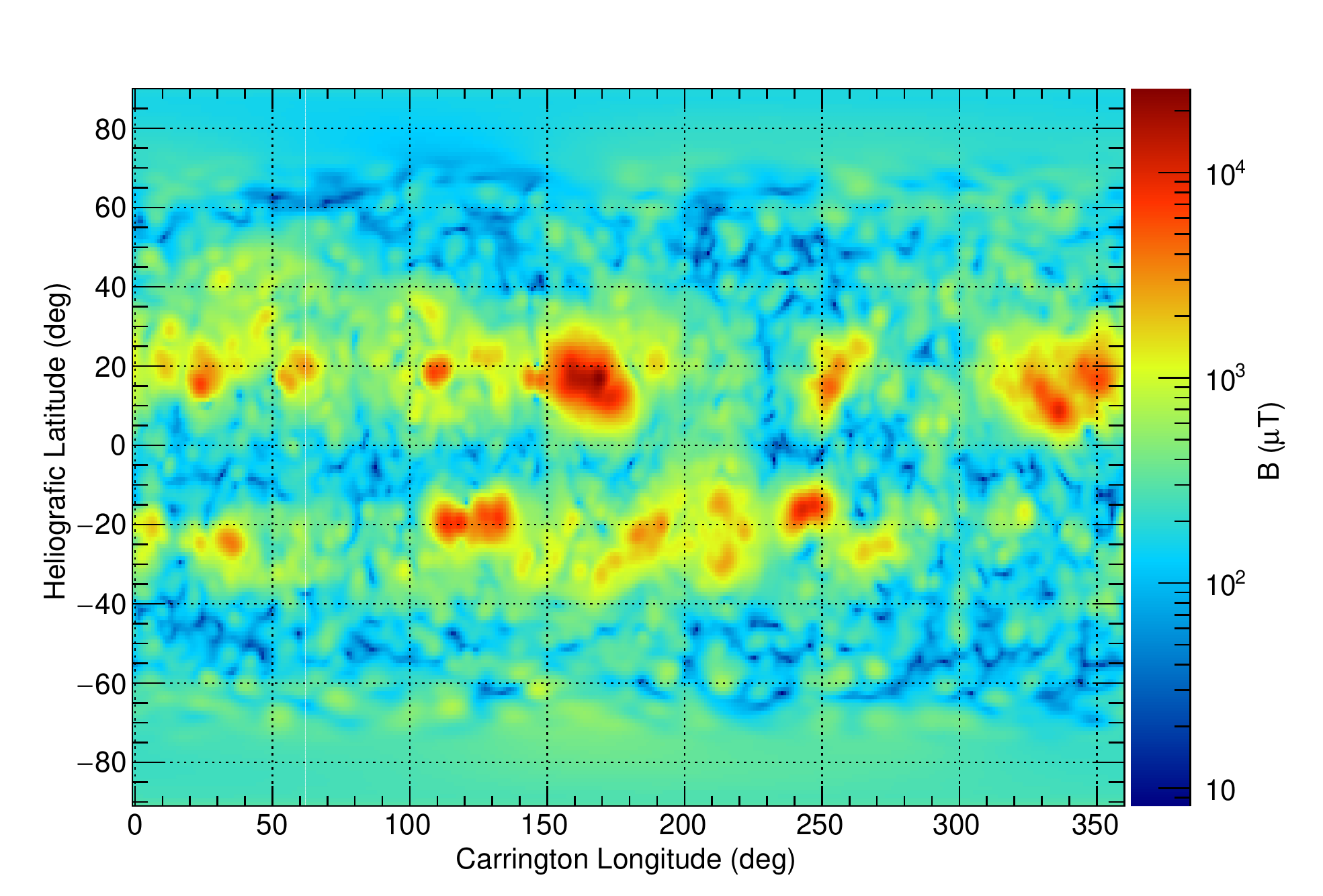}
\caption{Magnetic field intensity as a function of the Carrington longitude and latitude angles at $r=R_{\odot}$ for the CR 2111.}
\label{fig:bm0}
\end{figure}

\subsection{Inner magnetic field}
\label{sec:inmf}

The magnetic field near the Sun is complex and strongly time-dependent, and the coronal magnetic field is usually extrapolated from the observed photospheric fields. A widely adopted model is the potential field source surface (PFSS) model~\cite{Schatten1969,Hakamada1995}, in which the field is purely radial on a sphere of radius $R_{SS}$ (source surface). 
In our simulation we have implemented the field maps taken from the Solar Dynamics Observatory Joint Science Operations Center (JSOC)~\cite{Scherrer1995,jsoc}, which are calculated starting from the photospheric magnetic field observations~\cite{Titov_2008,Antiochos_2011,Sun2011} of the Helioseismic and Magnetic Imager (HMI)~\cite{hmi}, the Solar Dynamics Observatory (SDO)~\cite{sdo} and the Michelson Doppler Imager (MDI)~\cite{mdi} instrument on the Solar and Heliospheric Observatory (SOHO)~\cite{soho} and assuming $R_{SS}=2.5 R_{\odot}$. In each map the three components of the coronal magnetic field, $(B_r, B_\theta, B_\phi)$ are tabulated at 51 heights between the photosphere ($r=R_{\odot}$) and the source surface ($r=R_{SS}$). The field maps are available starting from the Carrington Rotation (CR) 2097 (May-June 2010).

In the present work we assume that the magnetic field inside the Sun is always equal to that at $r=R_{\odot}$. The intensity of the magnetic field on the solar surface is shown in Fig.~\ref{fig:bm0} for the CR 2111, covering the period from 2011-06-05 17h to 2011-07-03 00h. We point out that in small regions of the solar surface the field intensity can even exceed 10\units{G}. 

\subsection{Interplanetary magnetic field}
\label{sec:imf}

The interplanetary magnetic field (IMF) affects the propagation of cosmic rays in the solar system. In our simulation we describe the IMF using the Parker model~\cite{Parker:1958zz} for $r>R_{SS}$. The three components of the IMF are given by:

\begin{eqnarray}
B_{r} & = & \pm f B_{E} \left( \frac{R_{E}}{r} \right)^{2} \label{eq:br} \notag \\
B_{\theta} & = & 0 \\
B_{\phi} & = & -B_{r} \tan \xi \notag
\end{eqnarray}
The angle $\xi$ is defined as:

\begin{equation}
\tan \xi(r, \theta) = \frac{\omega_{S} \left( r - R_{SS} \right) \sin \theta}{v_{SW}}      
\end{equation}
where $\theta$ is the polar angle, $\omega_{S}=2.69 \times 10^{-6} \units{rad/s}$ is the angular velocity of the Sun (corresponding to a period of about $27 \units{days}$) and $v_{SW}$ is the velocity of the solar wind (its typical value is $400\units{km/s}$). At the distance $R_{SS}$ the components $B_{\phi}$ and $B_{\theta}$ are null, to ensure continuity with the PFSS model of the inner field (see Sect.~\ref{sec:inmf}).

In the previous equations the intensity of the field $B_E$ is given by $B_{E}=B_0/\sqrt{1+\tan^2 \xi(R_E,\pi/2)}$, where $B_{0}$ is the intensity of the magnetic field at the Earth (its typical value is about $5\units{nT}$), and $R_{E}=1\units{AU}$ is the Sun-Earth distance. The constant $f$ is given by:

\begin{equation}
f = 1 - 2H(\theta - \theta')    
\end{equation}
where $H$ is the Heaviside function and the angle $\theta'$ is the polar position of the heliospheric current sheet (HCS) defined as:

\begin{equation}
\theta' = \frac{\pi}{2} - \arctan \left[ 
\tan \alpha  \sin \left( \phi + \frac{\omega_{S} \left( r - R_{SS} \right)}{v_{SW}} \right)
\right]
\end{equation}
where $\phi$ is the azimuth angle and we have indicated with $\alpha$ the tilt angle, i.e. the maximum latitude of the HCS; finally, the $\pm$ sign in Eq.~\ref{eq:br} depends on the polarity of the magnetic field.

\begin{figure}[!htb]
\centering
\includegraphics[width=0.99\columnwidth,height=0.25\textheight,clip]{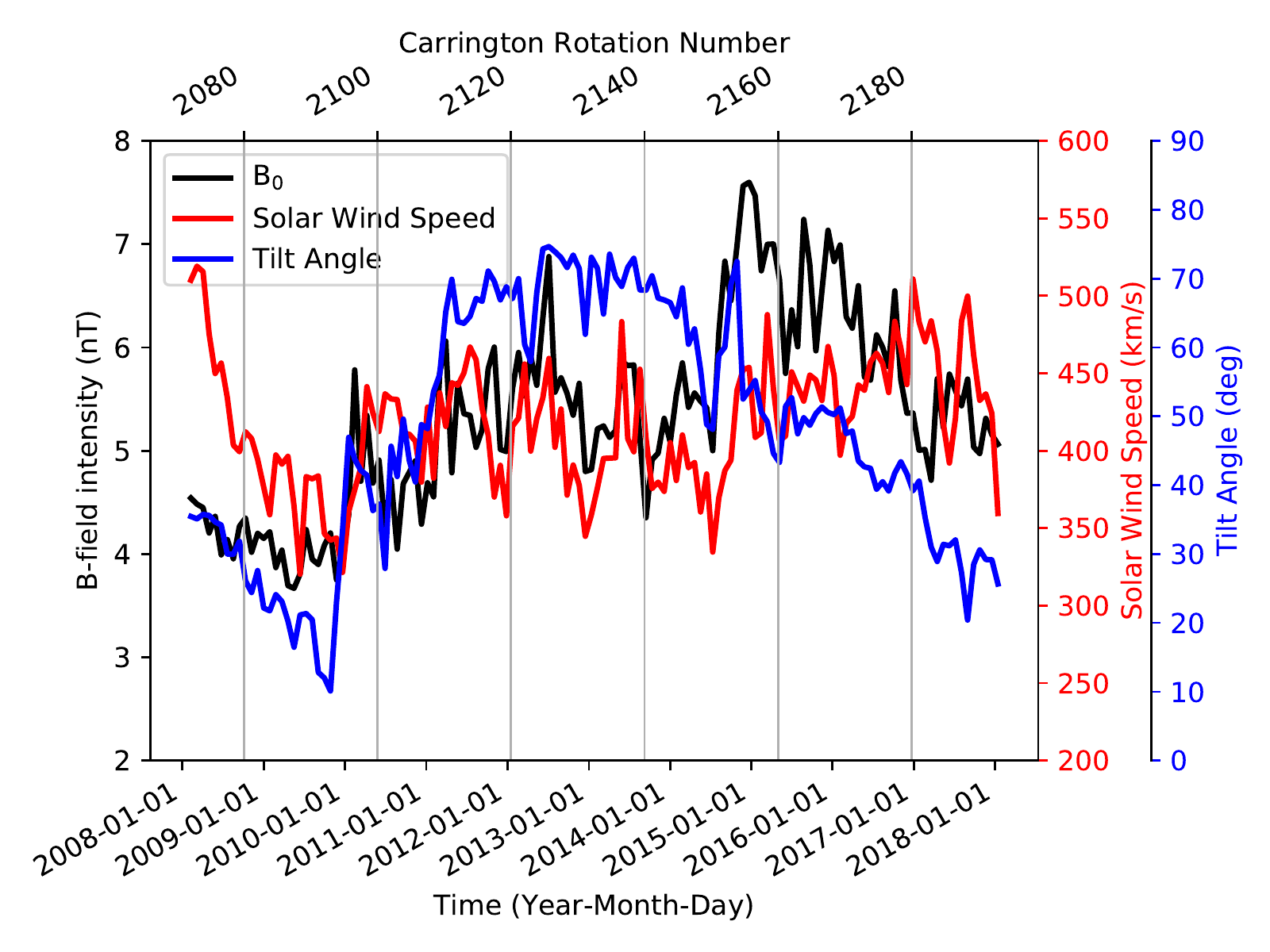}
\caption{Time evolution of $B_0$ (black line), of the solar wind speed $v_ {SW}$ (red line) and of the tilt angle $\alpha$ (blue line). The upper horizontal time scale shows the the CR numbers. The values of the tilt angle $\alpha$ and its polarity are taken from the Wilcox Solar Observatory public website~\cite{tiltweb}. The magnetic field at the Earth $B_{0}$ and the velocity of the solar wind $v_{SW}$ are taken from the observations of the ACE satellite extracted from the NASA/GSFC's OMNI dataset~\cite{omniweb,doi:10.1029/2004JA010649}.}
\label{fig:imf}
\end{figure}

Figure~\ref{fig:imf} shows the time evolution of the magnetic field $B_{0}$ at Earth, of the solar wind velocity $v_{SW}$ and of the tilt angle $\alpha$ averaged in each CR from 2008 to 2018.
The values of the tilt angle $\alpha$ and of its polarity are taken from the Wilcox Solar Observatory public website~\cite{tiltweb}, while the intensity of the magnetic field at the Earth $B_{0}$ and the velocity of the solar wind $v_{SW}$ are taken from the observations of the ACE satellite extracted from the NASA/GSFC's OMNI dataset~\cite{omniweb,doi:10.1029/2004JA010649}. 

In our simulation we have implemented the magnetic field configurations corresponding to a few CRs between 2011 and 2014, when the maximum of the solar cycle 24 occurred.

\section{Simulation results}
\label{sec:results}

\begin{figure}[!ht]
\centering
\includegraphics[width=0.99\columnwidth,height=0.23\textheight,clip]{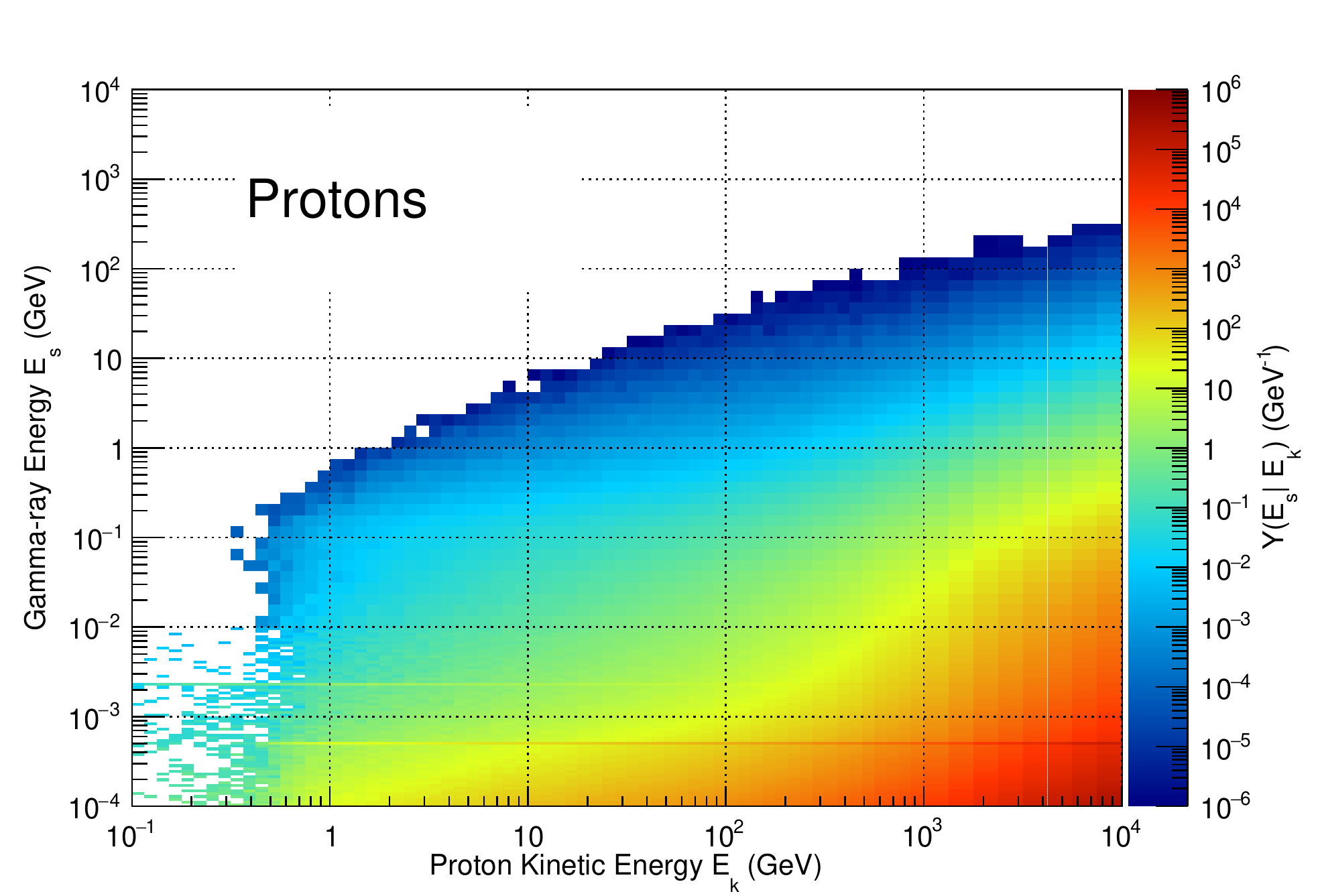}
\includegraphics[width=0.99\columnwidth,height=0.23\textheight,clip]{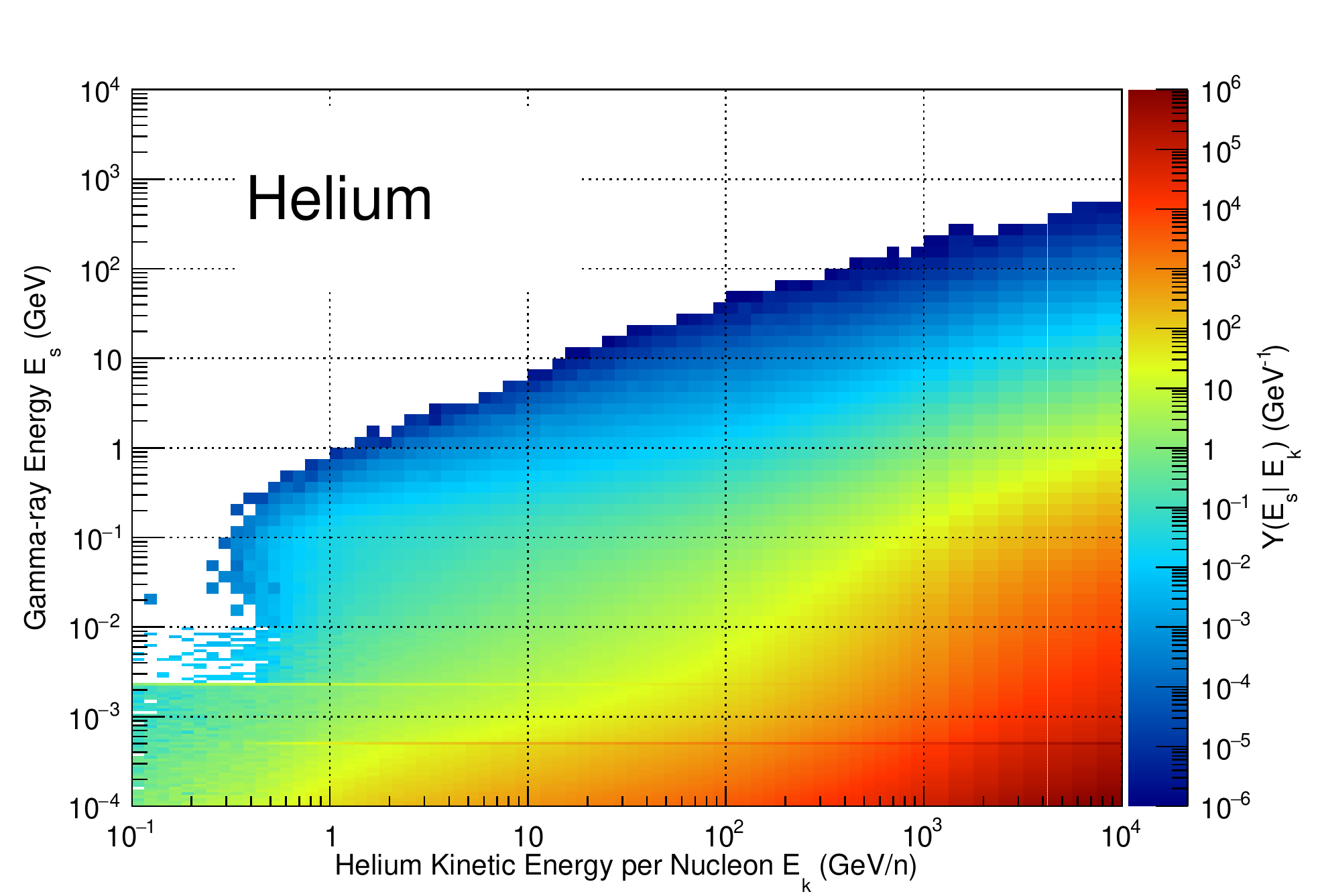}
\includegraphics[width=0.99\columnwidth,height=0.23\textheight,clip]{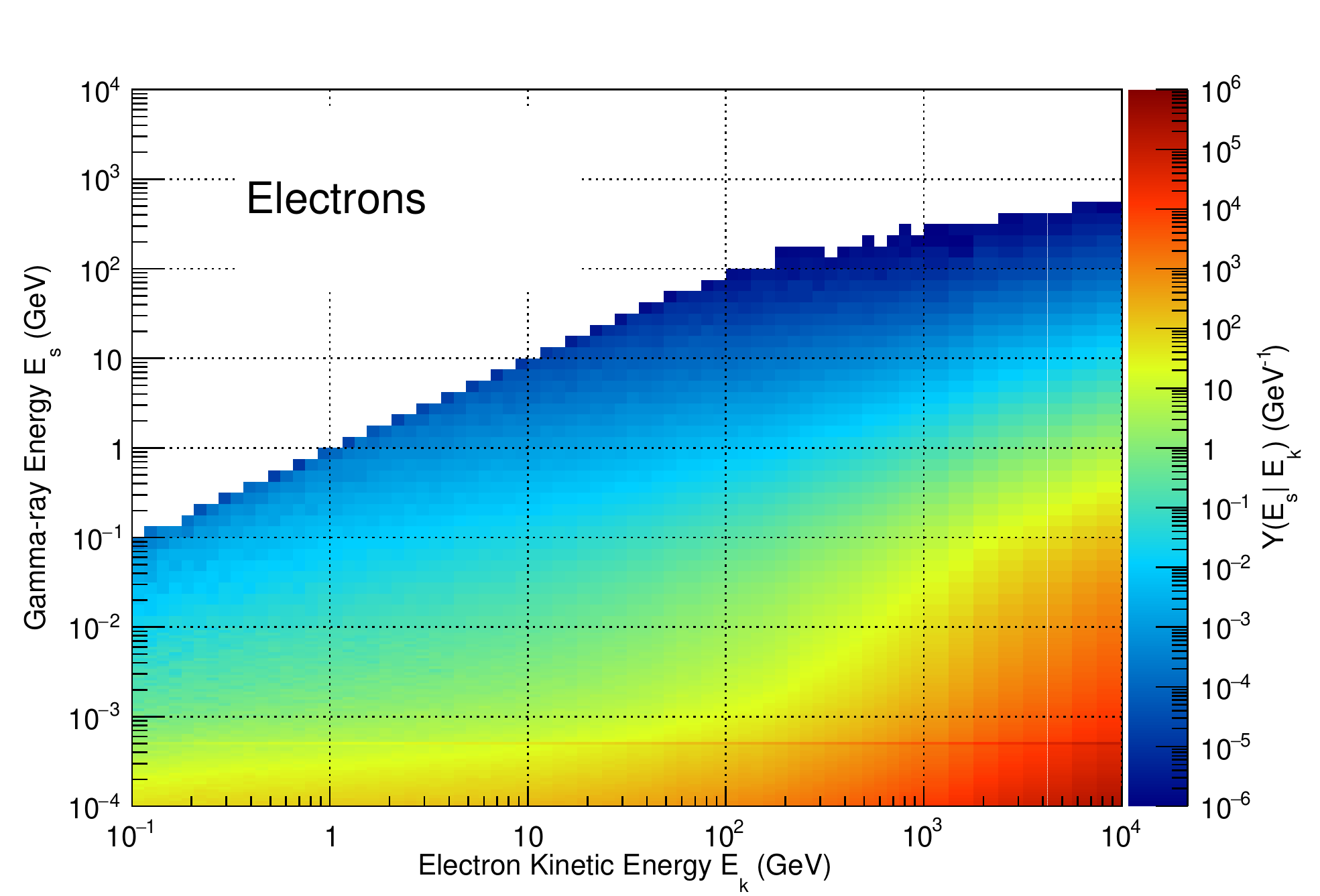}
\caption{Gamma-ray yields from protons (top panel), helium nuclei (middle panel) and electrons (bottom panel) as a function of the primary kinetic energy (or kinetic energy per nucleon in the case of helium primaries) (x-axis) and of the gamma-ray energy (y-axis). The color scale (z-axis) indicates the yields.}
\label{fig:yield}
\end{figure}

The yield of secondary particles produced from the $i$-th species of cosmic-ray primaries (here $i=p$, $e^-$ and $^{4}$He), $Y_{s,i}(E_{s} | E_k)$, is calculated by counting the secondary particles which escape from the generation surface. The yield is defined as:

\begin{equation}
Y_{s,i}(E_{s} |~ E_k) = \frac{N_{s,i}(E_{s} |~ E_k)}
{N_{i}(E_k) \Delta E_{s}}
\label{eq:yield}
\end{equation} 
where $N_{i}(E_k)$ is the number of primaries of the $i$-th species generated with kinetic energy $E_k$ ($E_k$ is expressed in units of\units{GeV} for primary electrons and protons and of\units{GeV/n} for primary nuclei) and $N_{s,i}(E_{s}|~E_k)$ is the number of secondaries of the species $s$ with energies between $E_{s}$ and $E_{s} + \Delta E_{s}$ produced by the primaries of the type $i$ with kinetic energy $E_k$ and escaping from the generation surface.
Fig.~\ref{fig:yield} shows the yields of gamma rays produced by protons (top panel), helium nuclei (middle panel) and electrons (bottom panel) as a function of the primary energy and of the gamma-ray energy. 

\begin{figure}[!ht]
\centering
\includegraphics[width=0.99\columnwidth,height=0.23\textheight,clip]{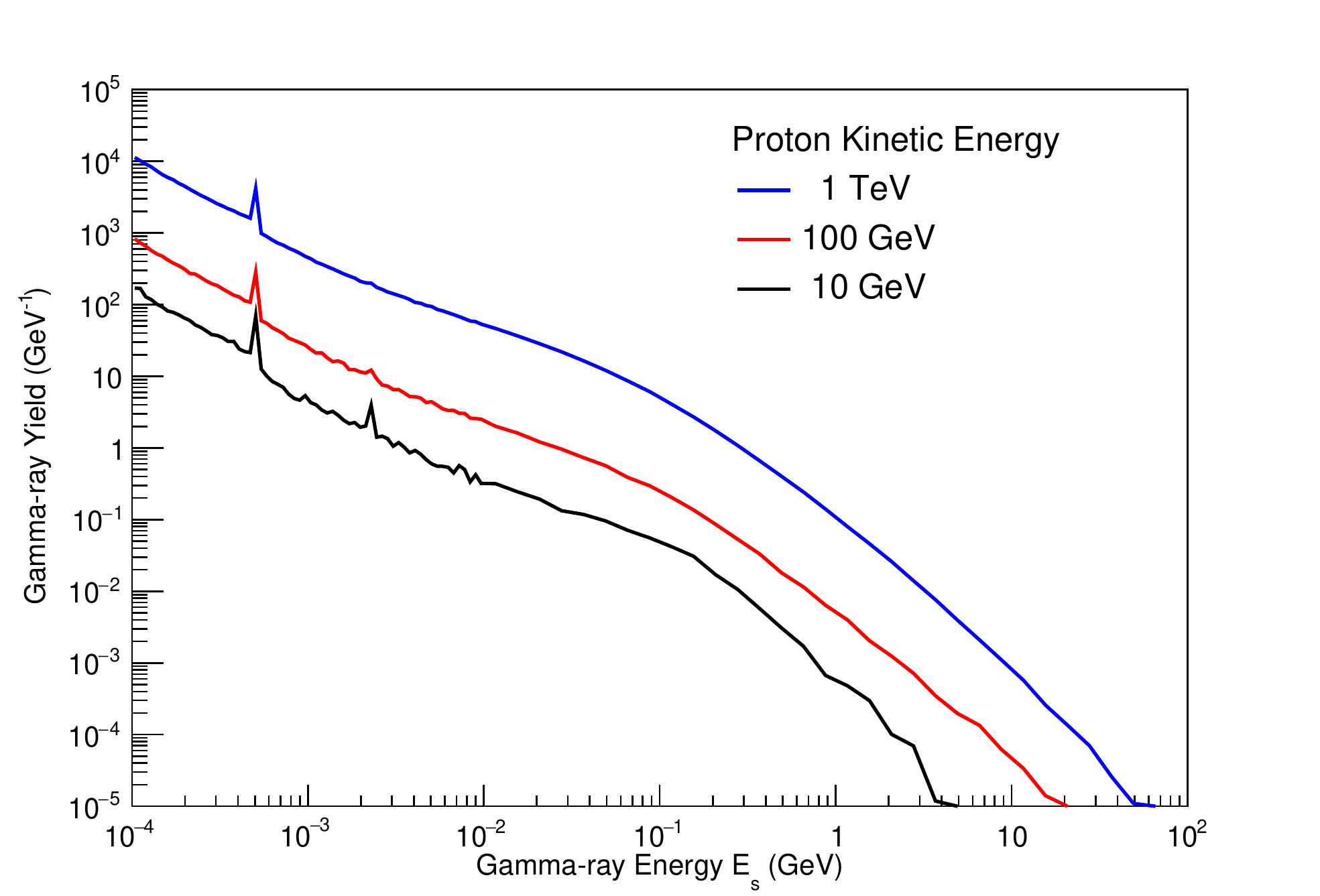}
\caption{Gamma-ray yields as a function of the gamma-ray energy for three different primary proton energies: 1\units{TeV} (blue line); 100\units{GeV} (red line); 10\units{GeV} (black line).}
\label{fig:pyieldpy}
\end{figure}

Fig.~\ref{fig:pyieldpy} shows the gamma-ray yields from primary protons for three different primary energies (10\units{GeV}, 100\units{GeV} and 1\units{TeV}). At fixed primary energy the yield roughly scales as $E_s^{-1}$ up to about 0.1$\units{GeV}$, while above this value it scales as $E_s^{-2}$. 
The soft component dominates the gamma-ray emission, and is mainly due to the secondary
production in the shower cascade for bremsstrahlung radiation effect.
This is the reason why the the average gamma-ray energy is much lower
than the energy of the parent particle.
The lines at $E_{s}=511\units{keV}$, corresponding to positron annihilation, are clearly visible for all primary energies. In the case of 10\units{GeV} primary protons, a line at $E_{s}=2.2\units{MeV}$ is also visible, corresponding to the neutron capture process, which tends to disappear as the primary proton energy increases.    

The differential intensity of secondary particles (in units of$\units{particles~GeV^{-1}~cm^{-2}~sr^{-1}~s^{-1}}$) emitted from the Sun is given by:
 
\begin{equation}
I_{s} (E_{s}) = \sum_{i} \int Y_{s,i}(E_{s} | E_k) ~ I_{i}(E_k)~dE_k
\label{eq:intensity}
\end{equation}
where $I_i(E_k)$ is the intensity of the $i$-th species of cosmic-ray primaries at the Sun. 

The flux of secondaries observed by a detector at Earth (in units of$\units{particles~GeV^{-1}~cm^{-2}~s^{-1}}$) is given by:

\begin{equation}
\label{eq:flux}
\phi_{s}(E_{s}) = \cfrac{\pi R_{SS}^{2}}{R_E^{2}}~I_{s}(E_s)~\mathcal{F}(E_s) 
\end{equation}
where $\mathcal{F}(E_s)$ is the fraction of secondaries with energy $E_s$ which are able to reach the Earth's orbit from the Sun. In our simulation we assume that the Earth's orbit lays on a sphere centered on the Sun with radius $r=R_E$. We point out here that not all secondaries emitted outwards from the Sun are able to reach the Earth. 
Charged particles are deflected by the IMF and, depending on their energy and initial direction, can be sent back to the Sun without reaching the Earth's orbit. In addition, there are some species of unstable secondaries, such as neutrons, which can decay during their journey from the Sun to the Earth.
In these cases, the fraction of secondaries reaching the Earth will be $\mathcal{F}(E_s) \leq 1$.
On the other hand, for the secondary gamma rays and neutrinos we assume $\mathcal{F}(E_s)=1$.\footnote{In our simulation we neglect the possible interactions of particles with the interplanetary dust.}

Cosmic rays impinging on the solar atmosphere are those which can reach the Sun from the interplanetary space. Hence the intensities $I_i(E_k)$ of the various cosmic-ray primaries in eq.~\ref{eq:intensity} are those at the surface of the generation sphere of radius $R_{SS}$, which differ from those measured at Earth, since not all cosmic rays reaching the Earth are able to continue their journey to the Sun. 

To evaluate the cosmic-ray intensities at the Sun we have used the custom code {\tt HelioProp}~\cite{Maccione:2012cu,Vittino:2017fuh}\footnote{See also \url{https://github.com/cosmicrays}.}, which describes the transport of cosmic rays in the solar system. We have simulated sets of pseudo-particles injected on the surface of the generation sphere with an isotropic and uniform distribution. The pseudo-particles are followed backwards in time during their propagation until they reach a sphere of radius $R_E$~\cite{Strauss_2011,Strauss2012,doi:10.1029/2007JA012280}. Their survival probabilities are used to scale the measured intensities of cosmic rays at Earth in order to properly set the intensities $I_i(E_k)$ in the right-hand side of eq.~\ref{eq:intensity}. 

In our simulations we assume that the intensity of cosmic rays measured at the Earth is the same across a sphere or radius $R_E=1\units{AU}$. At low energies ($<10\units{GeV}$) this assumption could be not valid because of a possible dependence on the charge sign of the propagation of cosmic rays from the outer space to $1\units{AU}$~\cite{Maccione:2012cu}. However, this effect is not expected to produce significant changes in our results, since only a small fraction of low-energy cosmic rays are able to reach the Sun.

We use the cosmic-ray intensities at Earth measured by AMS-02: the proton intensity is taken from Ref.~\cite{Aguilar:2015ooa}, the helium intensity is taken from Refs.~\cite{Aguilar:2015ctt,Aguilar:2017hno} and the electron~\footnote{We use the total intensity of electrons and positrons, and we refer to them as electrons.} intensity is taken from Ref.~\cite{Aguilar:2014fea}. We also use the AMS-02 spectra measured for different Bartels' rotations (BRs)~\cite{Aguilar:2018wmi,Aguilar:2018ons}.\footnote{A BR has a duration of exactly 27 days, close to the synodic CR of 27.2753 days. BR numbers start on 8 February 1832, while CR numbers start from November 9, 1853.} 

Since the AMS-02 spectra are available starting from about $0.4\units{GeV/n}$, we have extrapolated the data down to $0.1\units{GeV/n}$ by fitting the measured intensities with a function given by~\cite{Gaisser:2001jw}:

\begin{equation}
    I(E_k) = a ~ \left( E_k + b ~ e^{ -c ~ \sqrt{E_k} } \right)^{-\alpha}
\end{equation}

For the proton and helium we fit the data points up to the break energy around $200\units{GeV/n}$; then for larger energies we include a smooth break with a harder spectral index, as indicated by Refs.~\cite{Aguilar:2015ooa,Aguilar:2015ctt}. In the case of electrons we also take into account the DAMPE data~\cite{Ambrosi:2017wek} including a break at about $900\units{GeV}$. 

\begin{figure}[!ht]
\centering
\includegraphics[width=0.99\columnwidth,height=0.23\textheight,clip]{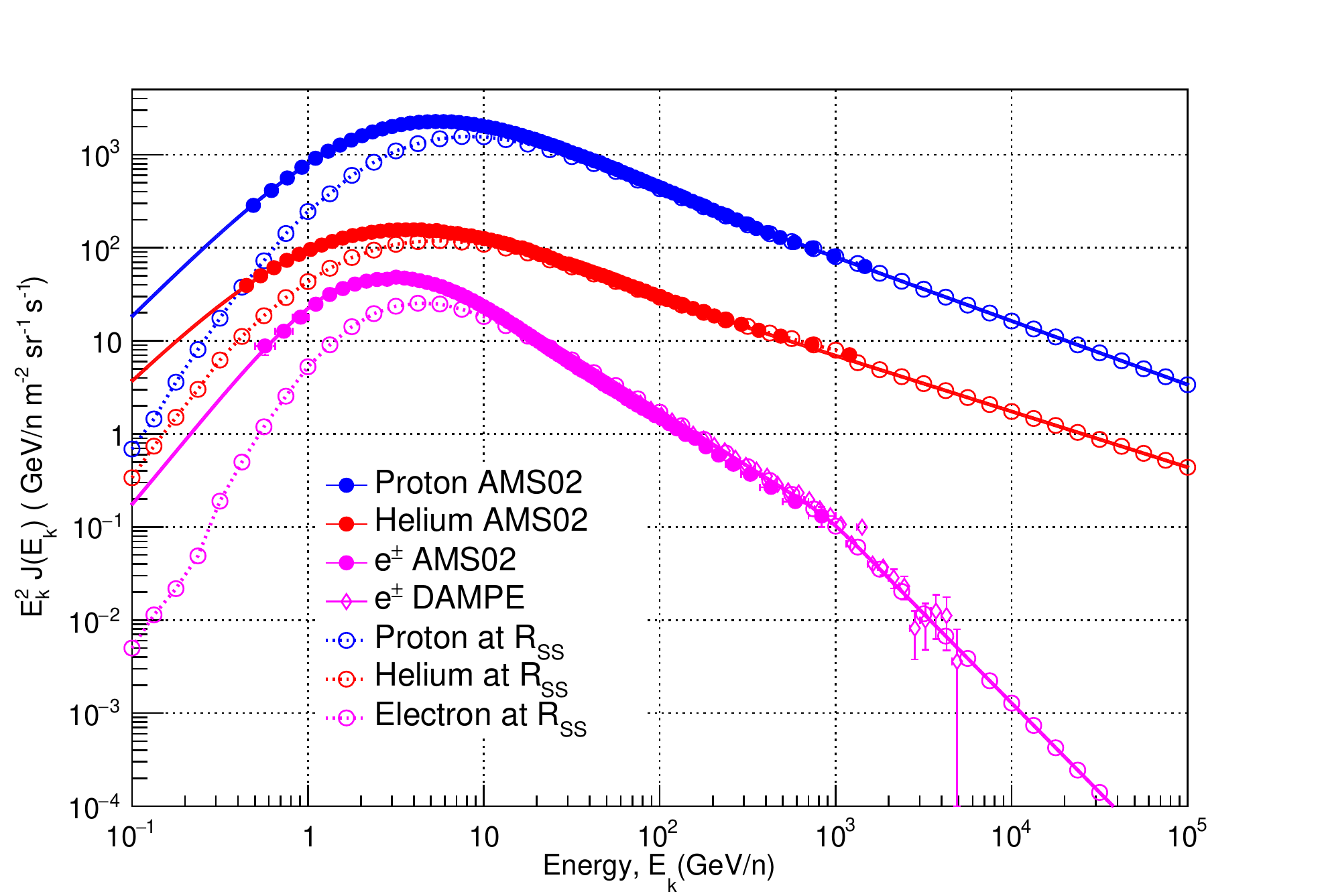}
\caption{Proton~\cite{Aguilar:2015ooa}, helium~\cite{Aguilar:2015ctt,Aguilar:2017hno} and electron intensities~\cite{Aguilar:2014fea,Ambrosi:2017wek} as function of the kinetic energy (kinetic energy per nucleon in the case of helium). The full circles correspond to the experimental data measured by AMS-02 at Earth during CR 2111; the solid lines indicate the results of the fits; the open circles with dashed lines indicate the intensities near the Sun at $R_{SS}=2.5 \times R_\odot$, evaluated by scaling the intensities at Earth with the survival probabilities obtained from {\tt Helioprop}~\cite{Maccione:2012cu}. }
\label{fig:cr}
\end{figure}

Figure~\ref{fig:cr} shows the results of the fitting procedure with the experimental data points corresponding to the CR 2111, covering the period from 2011, June $5^{th}$ to 2011, July $3^{rd}$. In Fig.~\ref{fig:cr} we also show the modulated spectra at the Sun, evaluated from those at the Earth with {\tt Helioprop}.

\begin{figure*}[!hbt]
\includegraphics[width=0.95\columnwidth,height=0.23\textheight]{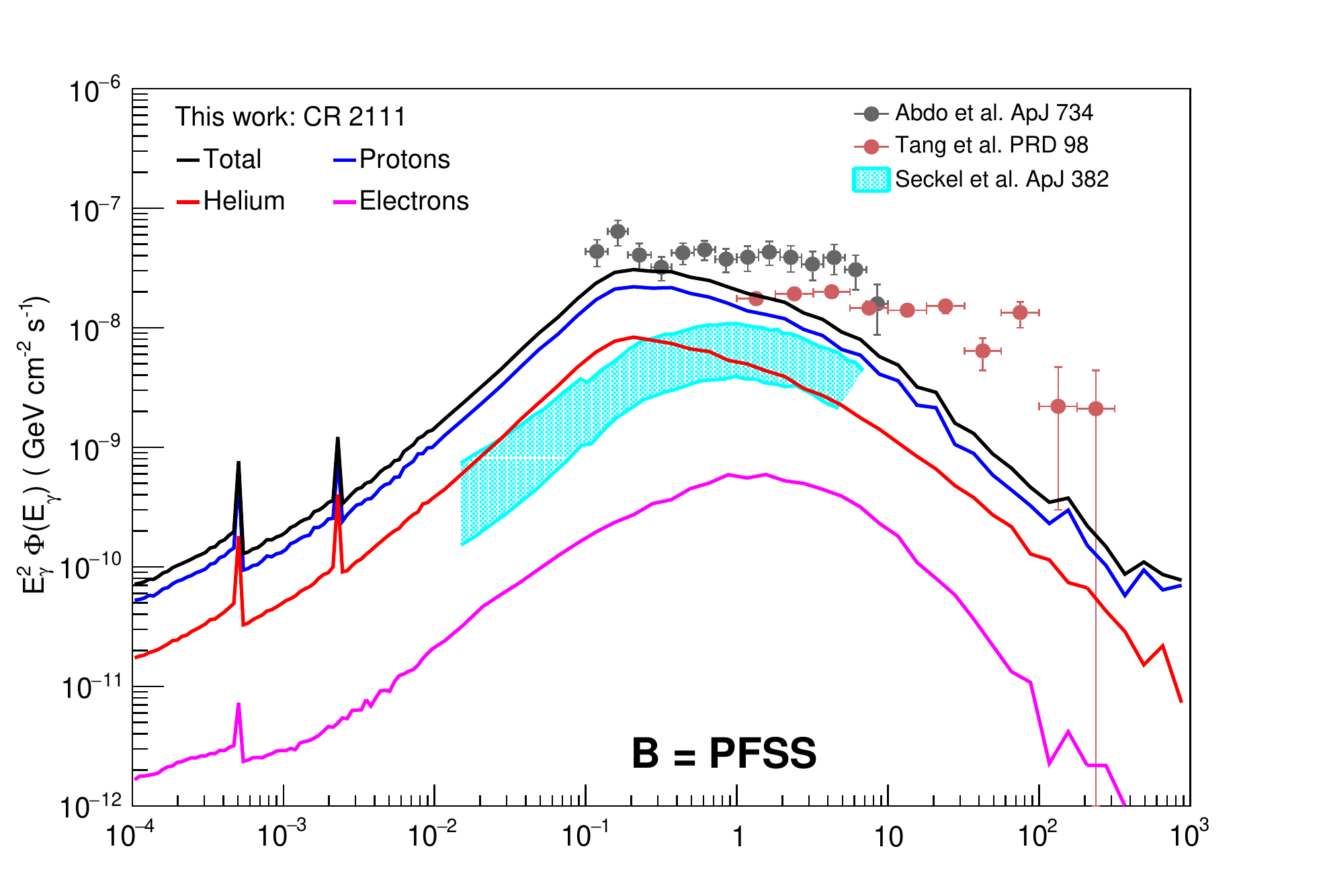}
\includegraphics[width=0.95\columnwidth,height=0.23\textheight]{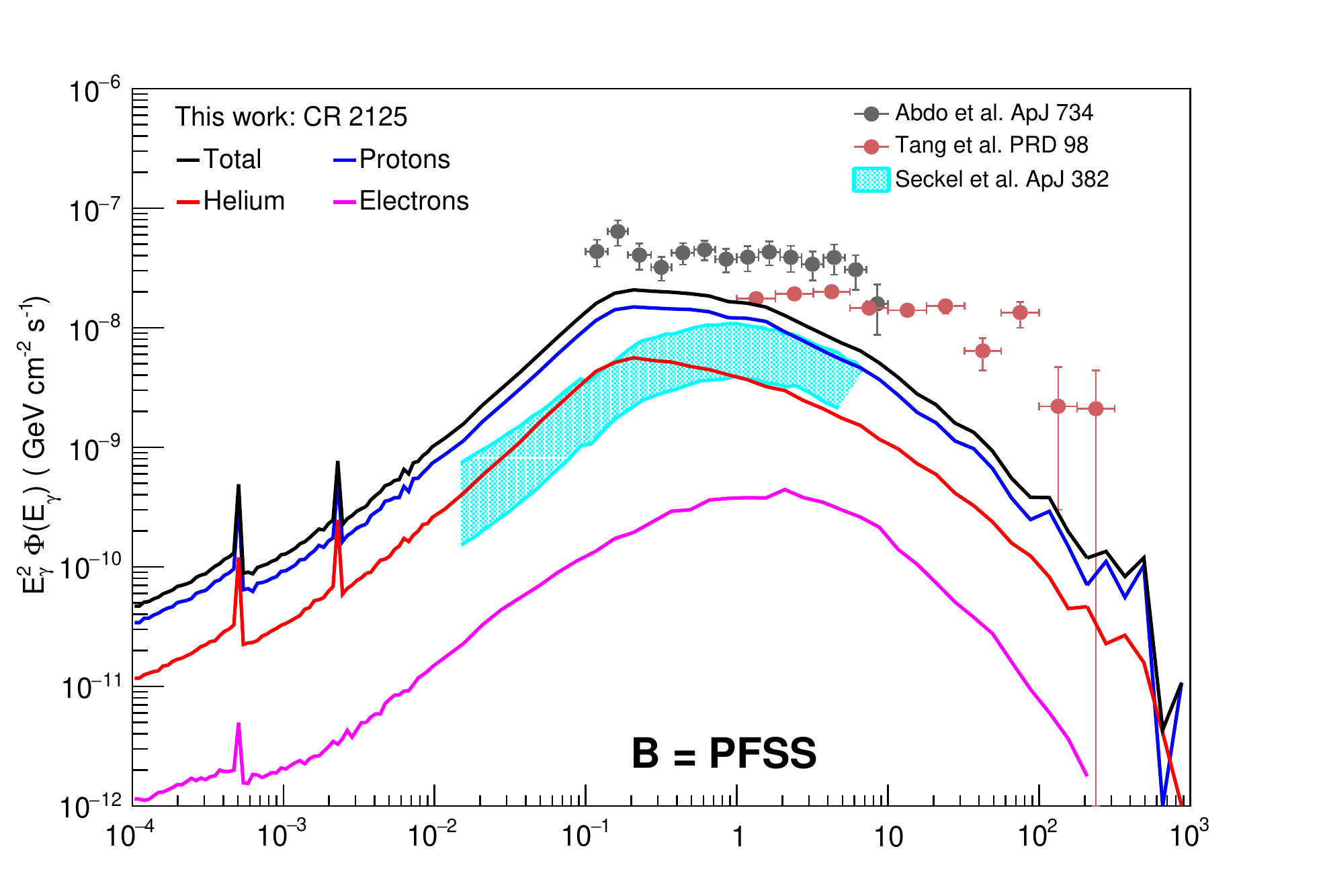}
\includegraphics[width=0.95\columnwidth,height=0.23\textheight]{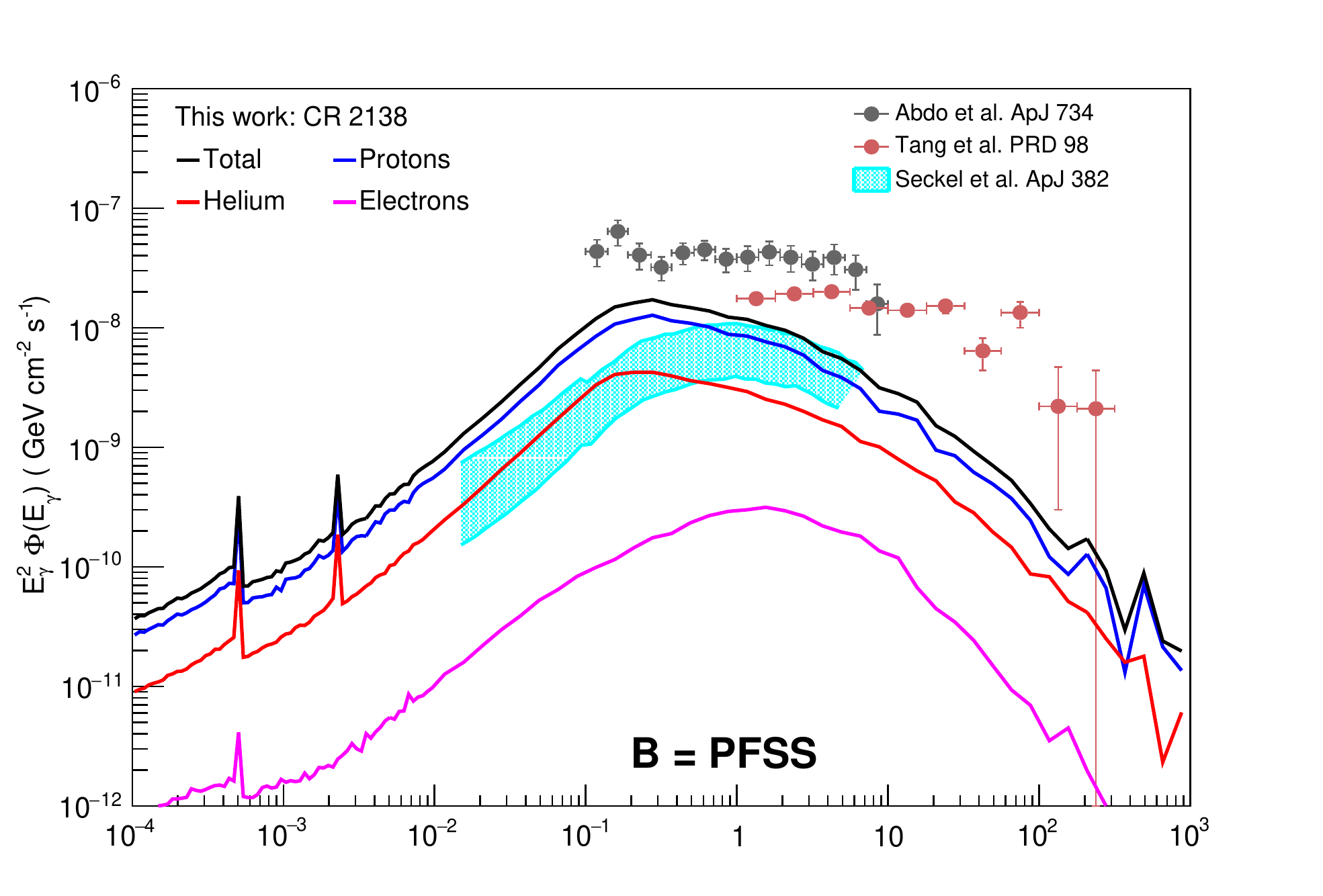}
\includegraphics[width=0.95\columnwidth,height=0.23\textheight]{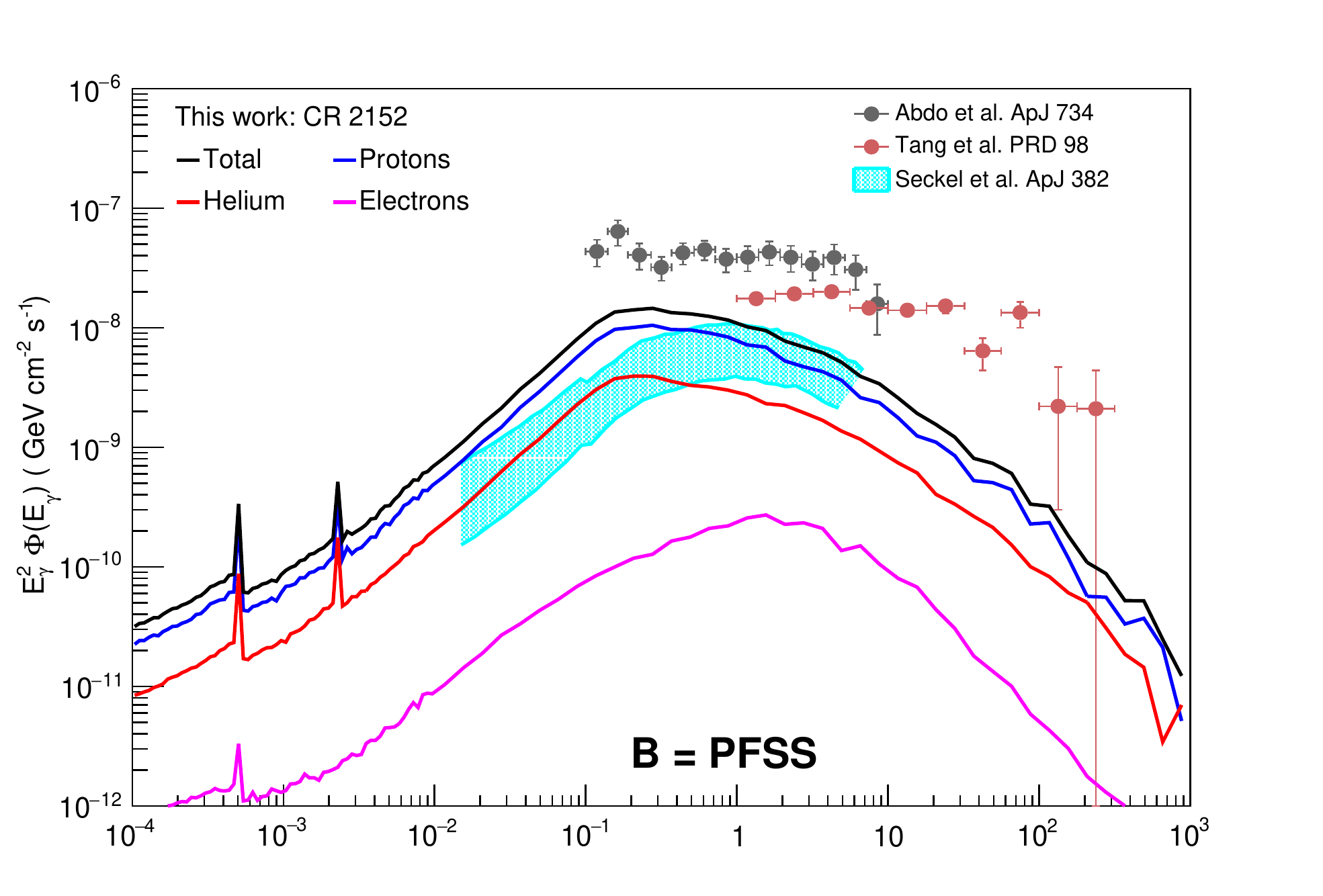}
\caption{Gamma-ray fluxes at the Earth for four different CRs. Top left panel:  CR 2111; top right panel:  CR 2125;  CR 2138; bottom right panel:  CR 2152. Black line: total emission; blue line: gamma rays from protons; red line: gamma rays from $^4He$ nuclei; magenta line: gamma rays from electrons. Light blue shadow region: Seckel et al. model~\cite{Seckel:1991ffa}; gray points: 1.5 years Fermi-LAT data~\cite{Abdo:2011xn}; dark red points: 9-years Fermi-LAT data~\cite{Tang:2018wqp}.}
\label{fig:c5s}
\end{figure*}

\begin{figure}[!ht]
    \centering
    \includegraphics[width=0.95\columnwidth,height=0.23\textheight]{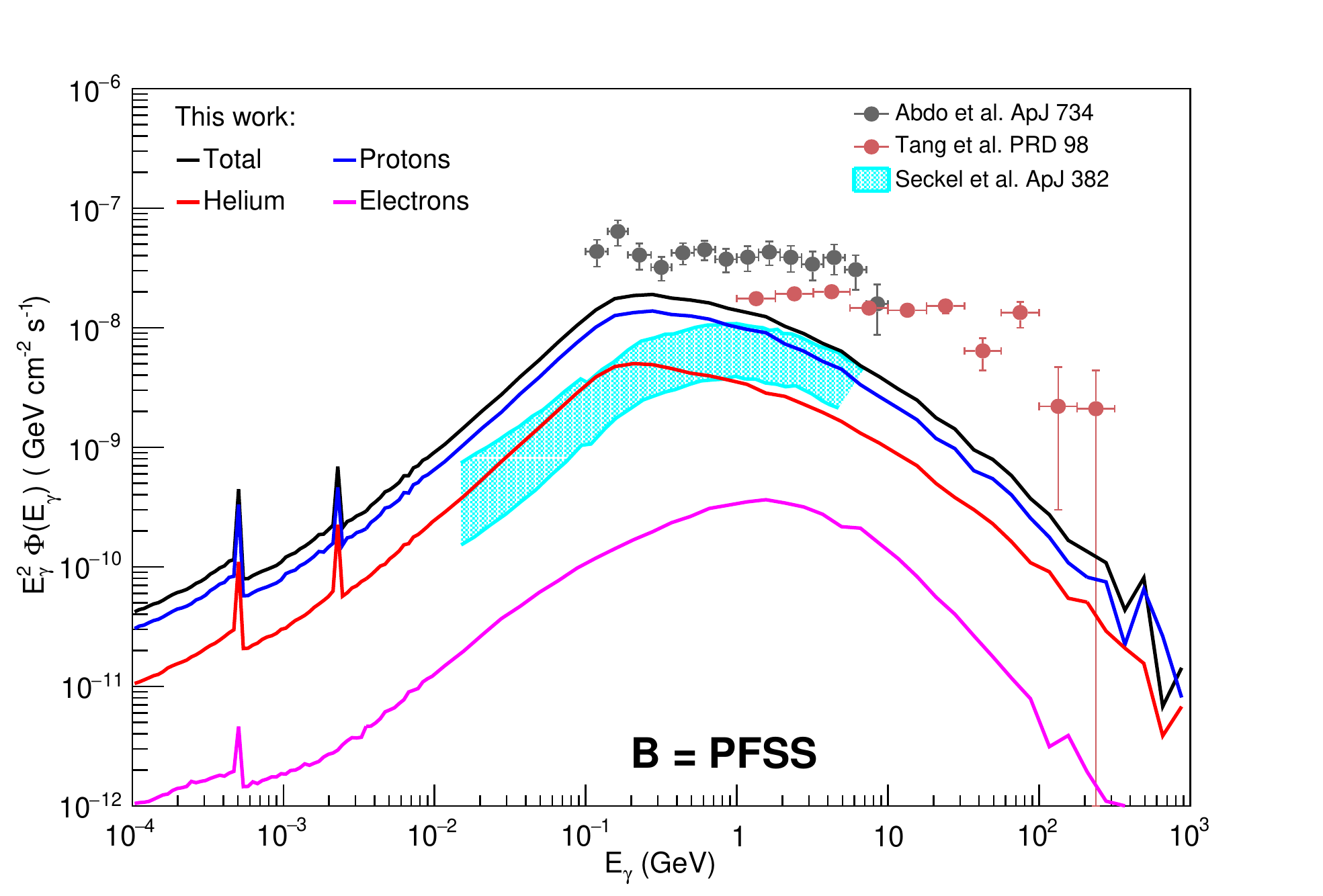}
    \caption{Gamma-ray flux at the Earth evaluated as the average of the fluxes of the four different Carrington rotations shown in Fig.~\ref{fig:c5s}. Color lines and data points have the same meanings as those in Fig.~\ref{fig:c5s}.}
    \label{fig:c5ave}
\end{figure}

Figure~\ref{fig:c5s} shows the gamma-ray fluxes at the Earth evaluated with our simulation set-up for four different CRs (2111, 2125, 2138 and 2152) spanning the period from June 2011 to June 2014, that covers the AMS-02 measurements. The calculated gamma-ray fluxes are slightly different at low energies ($< 1\units{GeV}$), due to the effect of the heliospheric magnetic field that affects both the cosmic-ray intensity at the Sun and the secondary yields. Finally, fig.~\ref{fig:c5ave} shows the gamma-ray flux at Earth obtained by averaging the fluxes calculated in the four different CRs.  

In Figs.~\ref{fig:c5s} and ~\ref{fig:c5ave} we show the total gamma-ray flux at Earth and the contributions of photons produced by the interaction of protons, Helium and electron primaries separately. The typical contributions of protons, helium and electron primaries to the total gamma-ray fluxes are of about 74\%, 24\% and 2\%, respectively.

The gamma-ray flux at the Earth exhibits two sharp peaks at $511\units{keV}$ and at about $2.2\units{MeV}$, due to the positron annihilation and to the neutron capture (in the hadronic interactions) respectively. These two lines could be used as reference to calibrate the low energy gamma-ray telescope proposed for the next decade, such as ASTROGAM~\cite{DeAngelis:2016slk,DeAngelis:2017gra} and AMEGO~\cite{mcenery2019allsky}. At energies above tens of\units{GeV} the calculated fluxes exhibit some fluctuations that are due to the limited statistic in the simulated data sets.\footnote{The simulation of high-energy primaries requires a high CPU consumption.}

In Figs.~\ref{fig:c5s} and~\ref{fig:c5ave} we also show the experimental results obtained with the Fermi-LAT~\cite{Atwood:2009ez} data for the disk component. The two LAT data sets correspond to a period of 1.5 years from August 2008 to January 2010, between the end of the $23^{rd}$ and the beginning of the $24^{th}$ solar cycle~\cite{Abdo:2011xn}, and to a period of 9 years from August 2008 to July 2017, spanning an almost full 11-years solar cycle~\cite{Tang:2018wqp}. We stress here that these measurements have been performed in different time windows from the one covered by our simulation. However, while our simulation predicts a peak in the spectral energy distribution of gamma rays at energies around $200\units{MeV}$, the data seem to indicate that the spectral energy distribution is almost flat up to beyond $10\units{GeV}$. This discrepancy could be due to the modeling of the inner magnetic field intensity, and will be further investigated in Sec.~\ref{sec:beff}. A possible cause of the discrepancy could be the modeling of the complex structure of the solar atmosphere. In addition, it could be due to the inverse Compton emission, that could produce high-energy gamma rays close to the solar surface that could be not well separated by the disc emission (we will discuss this point in Sec.\ref{sec:conc}).

As we discussed in Sec.~\ref{sec:inmf}, the inner PFSS field maps are available on the JSOC database only starting from CR 2097 (May-June 2010), and therefore they do not cover the whole time intervals of the analyses presented in Refs.~\cite{Abdo:2011xn,Tang:2018wqp}. In addition, the AMS-02 detector on the International Space Station started its operations only in May 2011 and, at present, their data are available until May 2017~\cite{Aguilar:2018ons,Aguilar:2018wmi}. Therefore, with the simulation set-up that we have implemented for this work, we are not able to make predictions on the gamma-ray flux in the period corresponding to the analysis of Ref.~\cite{Abdo:2011xn}. A detailed simulation of the whole time interval covered by the AMS-02 data would require a huge campaign, with a dedicated simulation for each CR in this period, but in any case it would not completely overlap with the time interval analyzed in Ref.~\cite{Tang:2018wqp}.   

In Figs.~\ref{fig:c5s} and~\ref{fig:c5ave} we also show the predictions of the gamma-ray flux at the Earth made by Seckel et al.~\cite{Seckel:1991ffa} under their nominal assumptions, taken from figure 7 in Ref.~\cite{Seckel:1991ffa}). The expected gamma-ray flux in each CR considered in the present work is always larger than the flux predicted in Ref.~\cite{Seckel:1991ffa}. The differences can be due to the different models used for describing the solar atmosphere and the inner magnetic field and to the different approach used in the simulation. In fact, while the authors of Ref.~\cite{Seckel:1991ffa} have evaluated the gamma-ray flux with a semi-analytical calculation with a simplified geometry, we have implemented a full Monte Carlo simulation with the complete geometry of the Sun.

\begin{figure*}[!htb]
    \centering
\includegraphics[width=0.95\columnwidth,height=0.23\textheight]{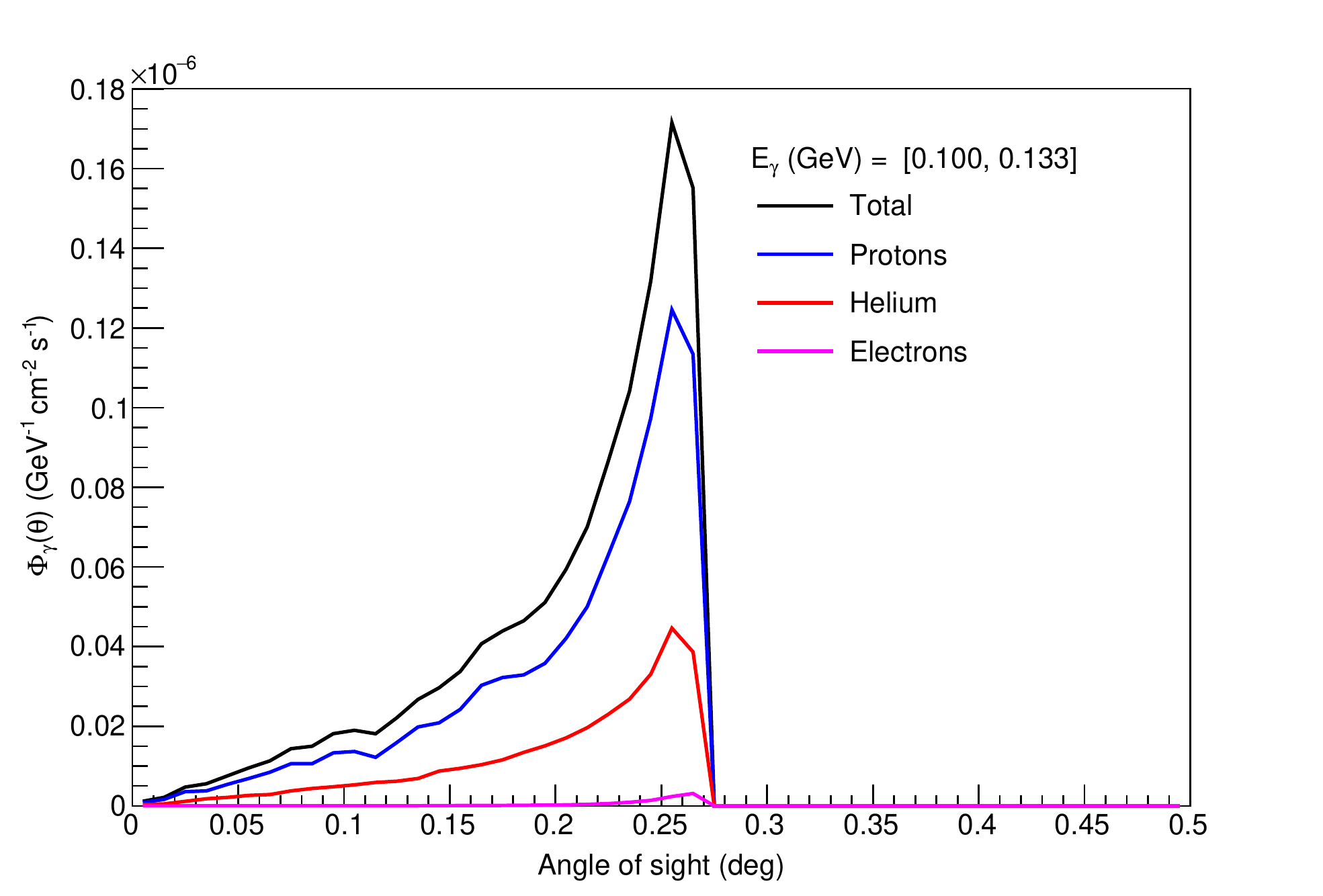}
\includegraphics[width=0.9\columnwidth,height=0.23\textheight]{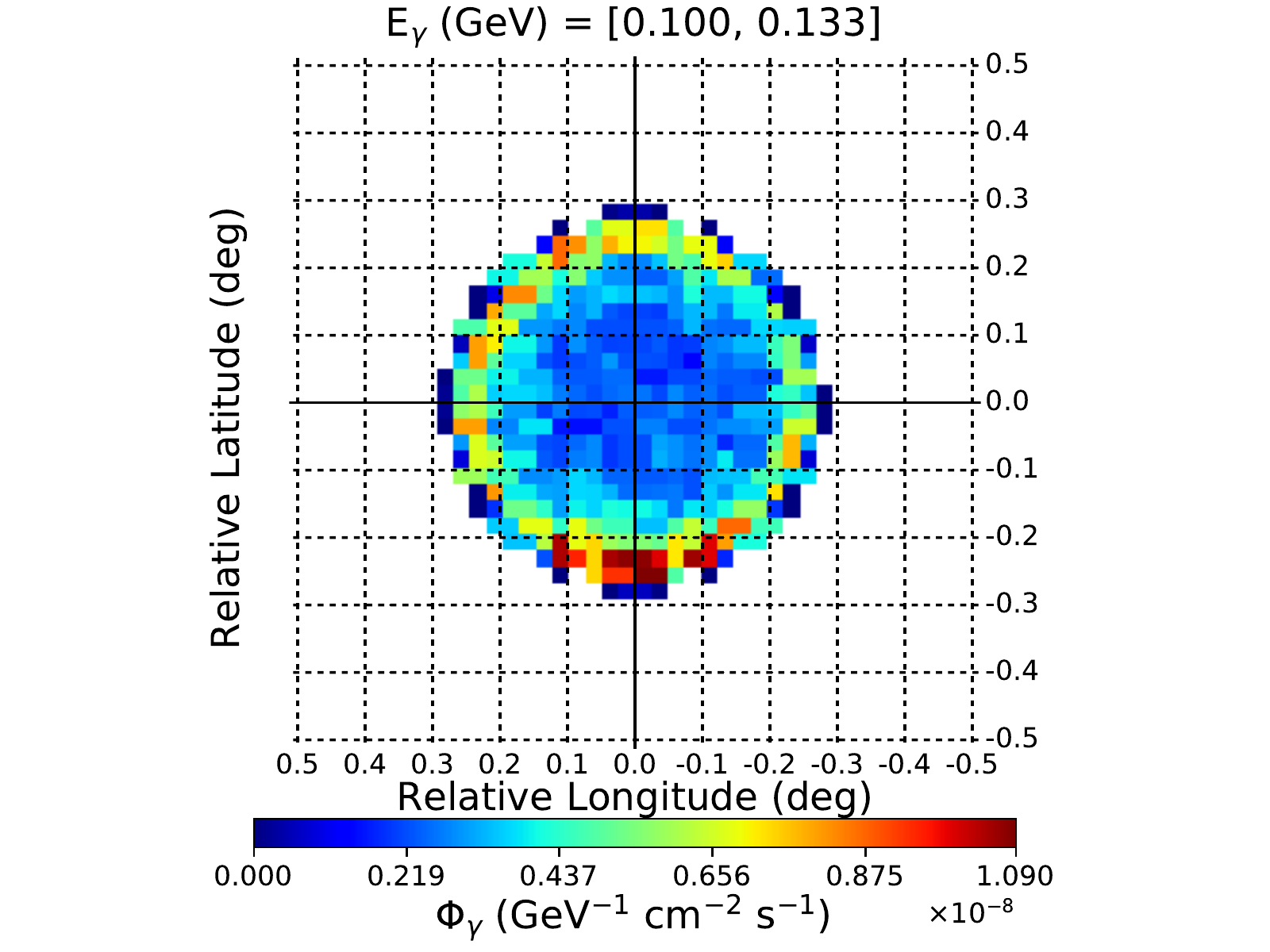}
\includegraphics[width=0.95\columnwidth,height=0.23\textheight]{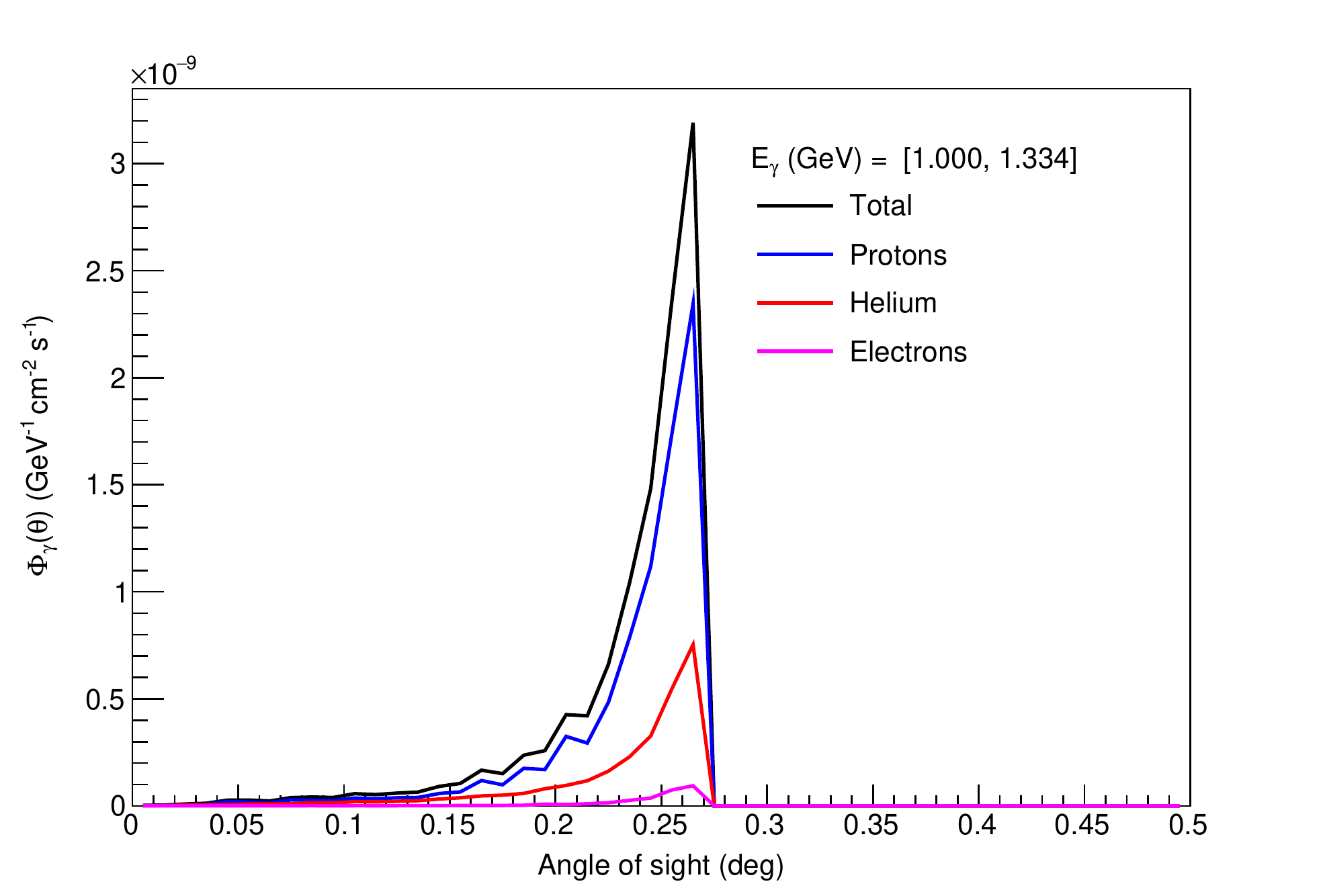}
\includegraphics[width=0.9\columnwidth,height=0.23\textheight]{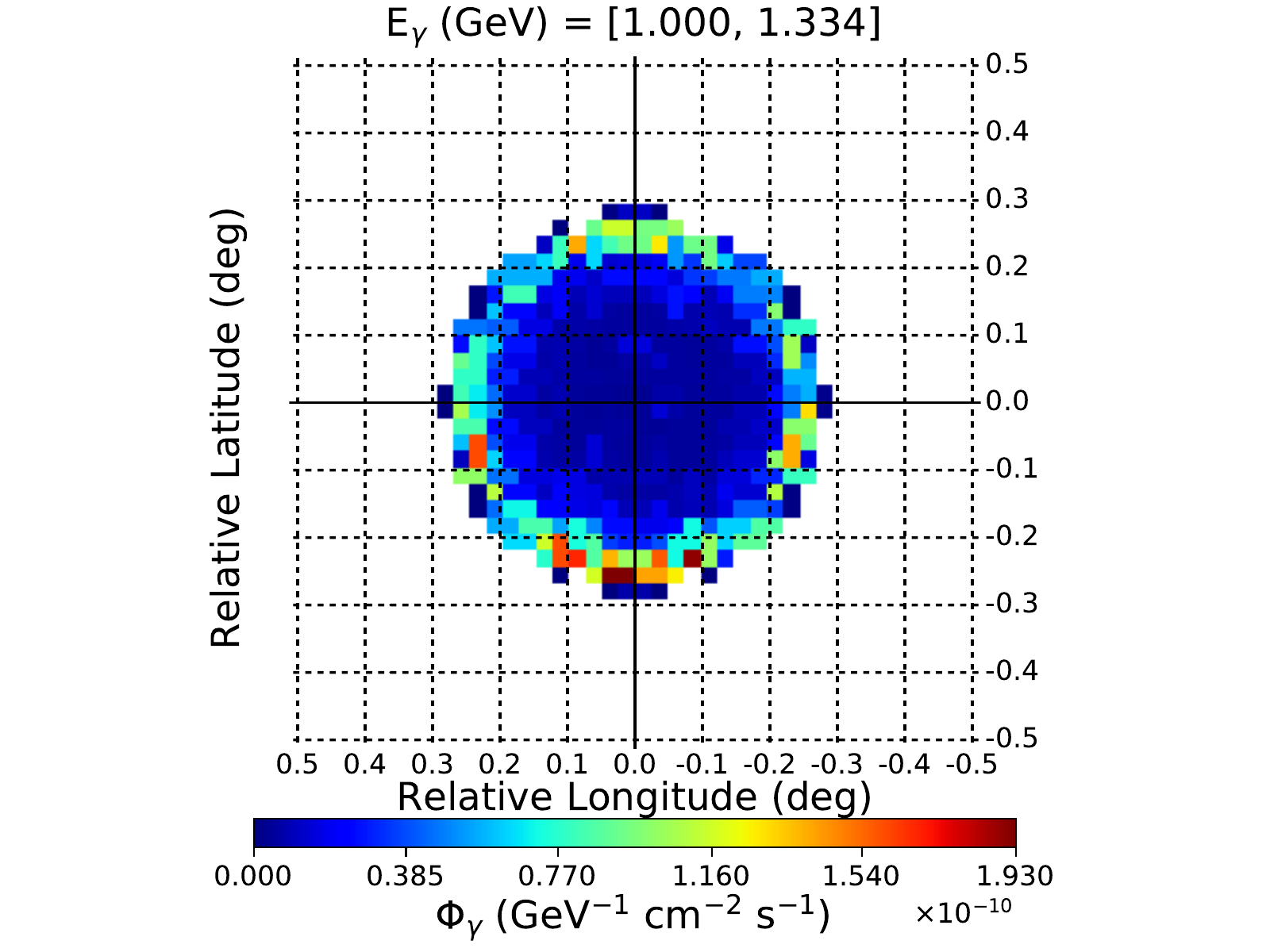}
\caption{Left panels: gamma-ray fluxes at the Earth as a function of the angle of sight. Right panels: spatial map of the solar gamma-ray emission built with the HEALPix pixelization with $N_{side}=2048$. Two different energy bins are considered: $[0.1, 0.133]\units{GeV}$ (top row)  and $[1, 1.33]\units{GeV}$ (bottom row).}
\label{fig:ang}
\end{figure*}

We have also cross-checked our results by back-propagating from the Sun to the Earth each particle simulated with {\tt FLUKA}. Given a cosmic-ray primary at the Sun, a particle with opposite charge and with opposite direction is back-propagated from the generation sphere of radius $R_{SS}$ to the sphere of radius $R_E$. If this particle is able to reach the Earth, the primary particle assigned a survival probability $P_{surv}=1$, otherwise it is assigned $P_{surv}=0$. With this procedure, the secondary yield can be calculated as:

\begin{equation}
Y_{s,i}(E_{s} |~ E_{k}) = \frac{N_{s,i}(E_{s} |~ E_{k},~P_{surv}=1)}
{N_{i}(E_{k}) \Delta E_{s}}
\label{eq:yield1}
\end{equation} 
where $N_{i}(E_{k})$ is the number of primaries of the $i$-th species generated with kinetic energy $E_{k}$ and $N_{s,i}(E_{s}|~E_{k},~P_{surv}=1)$ is the number of secondaries of the $i$-th species with energies between $E_{s}$ and $E_{s} + \Delta E_{s}$ produced by the primaries of the type $i$ with kinetic energy $E_k$ and $P_{surv}=1$. In this way, the secondary spectra at the Earth can be calculated inserting in the right-hand side of eq.~\ref{eq:intensity} the proton, helium and electron intensities measured at the Earth. Using this procedure we find the same results as when we evaluate the intensities of primary CRs at the Sun with {\tt HelioProp}.

In Tab.\ref{tab:IntFlux} we show the integral fluxes of gamma rays at Earth above 100\units{MeV}, 1\units{GeV} and 10\units{GeV} respectively, for the four CRs considered in this section. The integral flux decreases with increasing CR number, as the Sun approaches to its maximum activity.

The secondary productions at high energies occur close the solar surface, where the secondary are emitted in a low-density medium in the forward direction with respect to the high-energy primary particles. However, the combination of the solar magnetic field with the solar atmosphere density profile can affect the emission, even at high energies. Figure~\ref{fig:ang} shows the gamma-ray flux seen at the Earth as a function of the angle of sight for two different energy bins, i.e. $[0.1, 0.133] \units{GeV}$ and $[1, 1.33]\units{GeV}$. We also show the corresponding spatial emission maps centered on the Sun and built with the HEALPix pixelization~\cite{Gorski_2005}~\footnote{HEALPix website – currently \url{http://healpix.sourceforge.net} or \url{https://healpix.sourceforge.io.}}. The emission is mainly located nearby the solar surface and it becomes much narrow at higher energies.   

\begin{table}[!t]
    \centering
    \small
    \begin{tabular}{||c|c|c|c||} \hline
    CR & $\Phi_{\gamma}(>100\units{MeV})$ & $\Phi_{\gamma}(>1\units{GeV})$ & $\Phi_{\gamma}(>10\units{GeV})$ \\
    & $\times 10^{-7}\units{cm^{-2}~s^{-1}}$ & $\times 10^{-8}\units{cm^{-2}~s^{-1}}$ & $\times 10^{-9}\units{cm^{-2}~s^{-1}}$ \\ \hline
     2111 & 2.59 $\pm$ 0.02 & 1.42 $\pm$ 0.02 & 2.61 $\pm$ 0.10  \\ \hline
     2125 & 1.79 $\pm$ 0.01 & 1.16 $\pm$ 0.02 & 2.19 $\pm$ 0.08 \\ \hline
     2138 & 1.38 $\pm$ 0.01 & 0.84 $\pm$ 0.02 & 1.66 $\pm$ 0.06  \\ \hline
     2152 & 1.23 $\pm$ 0.01 & 0.74 $\pm$ 0.02 & 1.51 $\pm$ 0.05  \\ \hline
    \end{tabular}
    \caption{Gamma-ray fluxes at  Earth above $100\units{MeV}$, $1\units{GeV}$ and $10\units{GeV}$ respectively, in four different Carrington rotations.}
    \label{tab:IntFlux}
\end{table}

\begin{figure}[!ht]
    \centering
    \includegraphics[width=0.95\columnwidth,height=0.23\textheight]{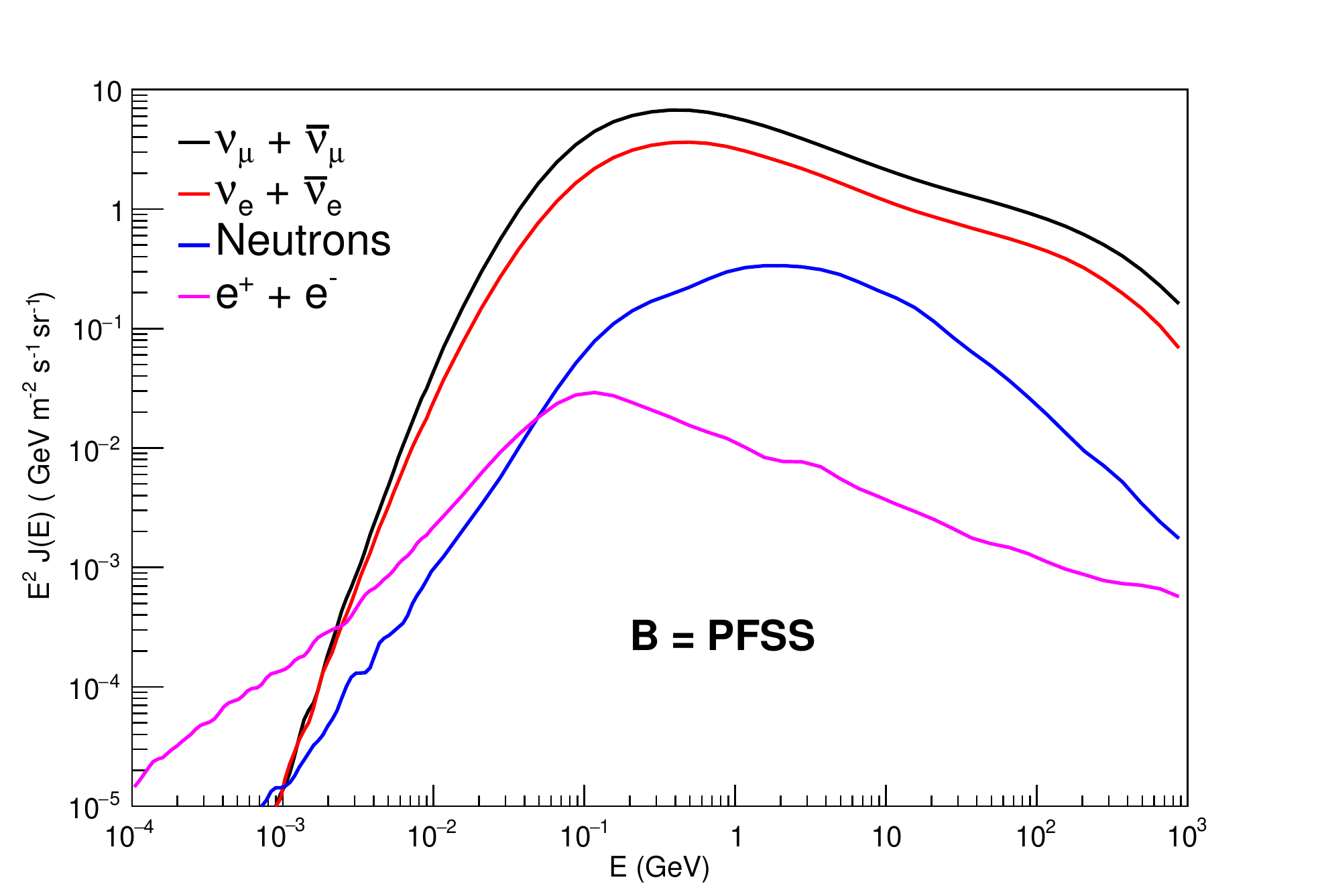}
    \caption{Average intensity at the Sun of $\nu_\mu + \bar{\nu}_\mu$ (black line), $\nu_e + \bar{\nu}_e$ (red line), neutrons (blue line),  $e^+ + e^-$ (magenta line) for CRs 2111, 2125, 2138 and 2152.}
    \label{fig:c6ave}
\end{figure}

Figure~\ref{fig:c6ave} shows the intensity of different species of secondaries (muon neutrinos and antineutrinos, electron neutrinos and antineutrinos, neutrons, electrons and positrons)  produced by the interactions of cosmic rays with the Sun, evaluated on the generation sphere. The values of the intensities are obtained by averaging the results in the four CRs mentioned above. To calculate the fluxes at the Earth, the decays of unstable particles during their journey from the Sun to the Earth should be taken into account, as well as the propagation of charged particles in the IMF. We also remark here that in the calculation of the neutrino and antineutrino fluxes we did not include their interactions in the Sun (their absorption is negligible below 10\units{GeV}) and their possible oscillations~\cite{Edsjo:2017kjk}.
The expected neutrino intensity is similar to that calculated in Ref.~\cite{Edsjo:2017kjk}, and above 100\units{GeV} it is higher than the intensity of neutrinos produced in cosmic-ray showers in the Earth's atmosphere  (see for example~\cite{Illana:2010gh}).

\begin{figure*}[!ht]
    \centering
    \includegraphics[width=0.95\columnwidth,height=0.23\textheight]{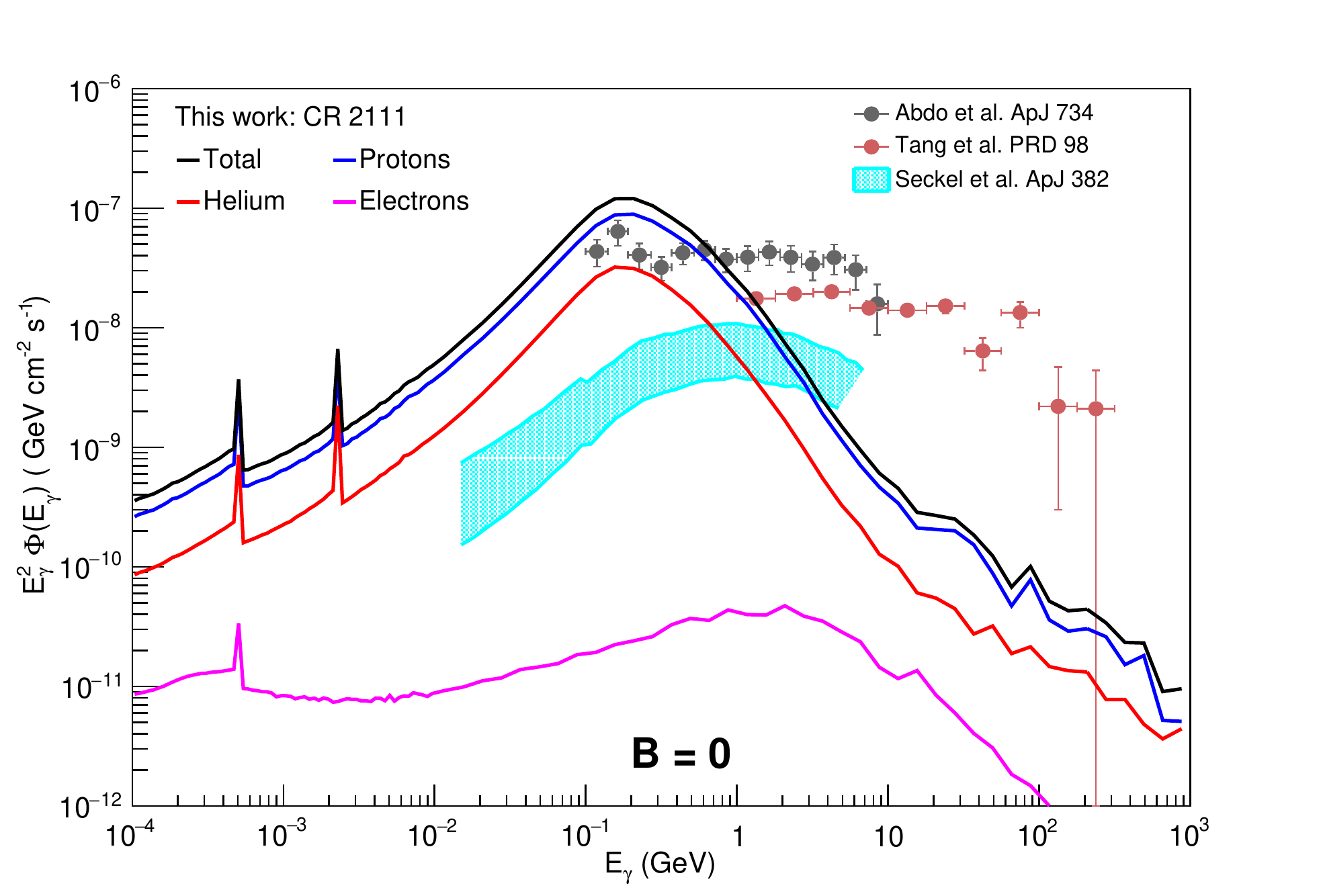}
    \includegraphics[width=0.95\columnwidth,height=0.23\textheight]{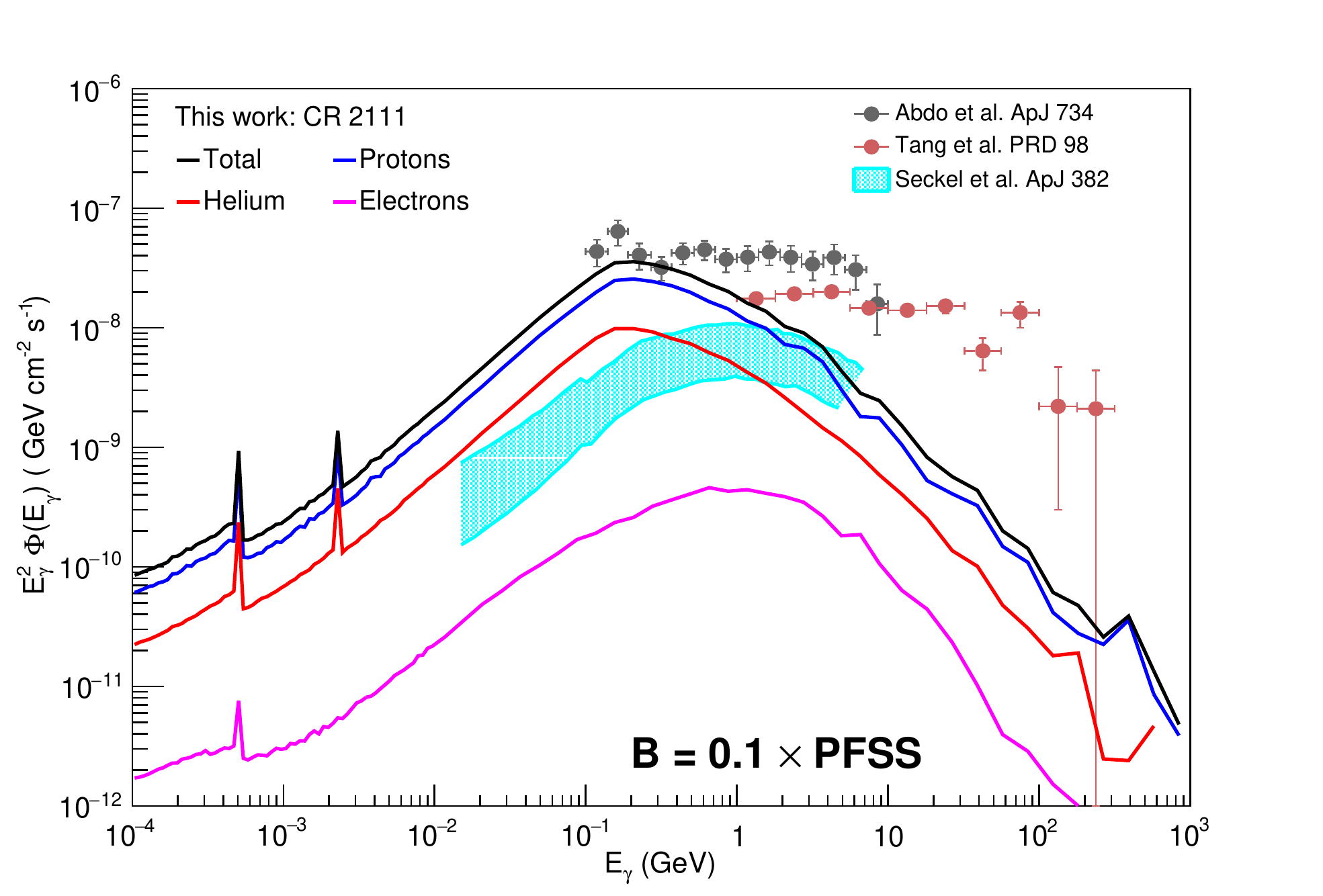}
    \includegraphics[width=0.95\columnwidth,height=0.23\textheight]{c5_gamma_ModS4_gs98_CR_2111_0_4.pdf}
    \includegraphics[width=0.95\columnwidth,height=0.23\textheight]{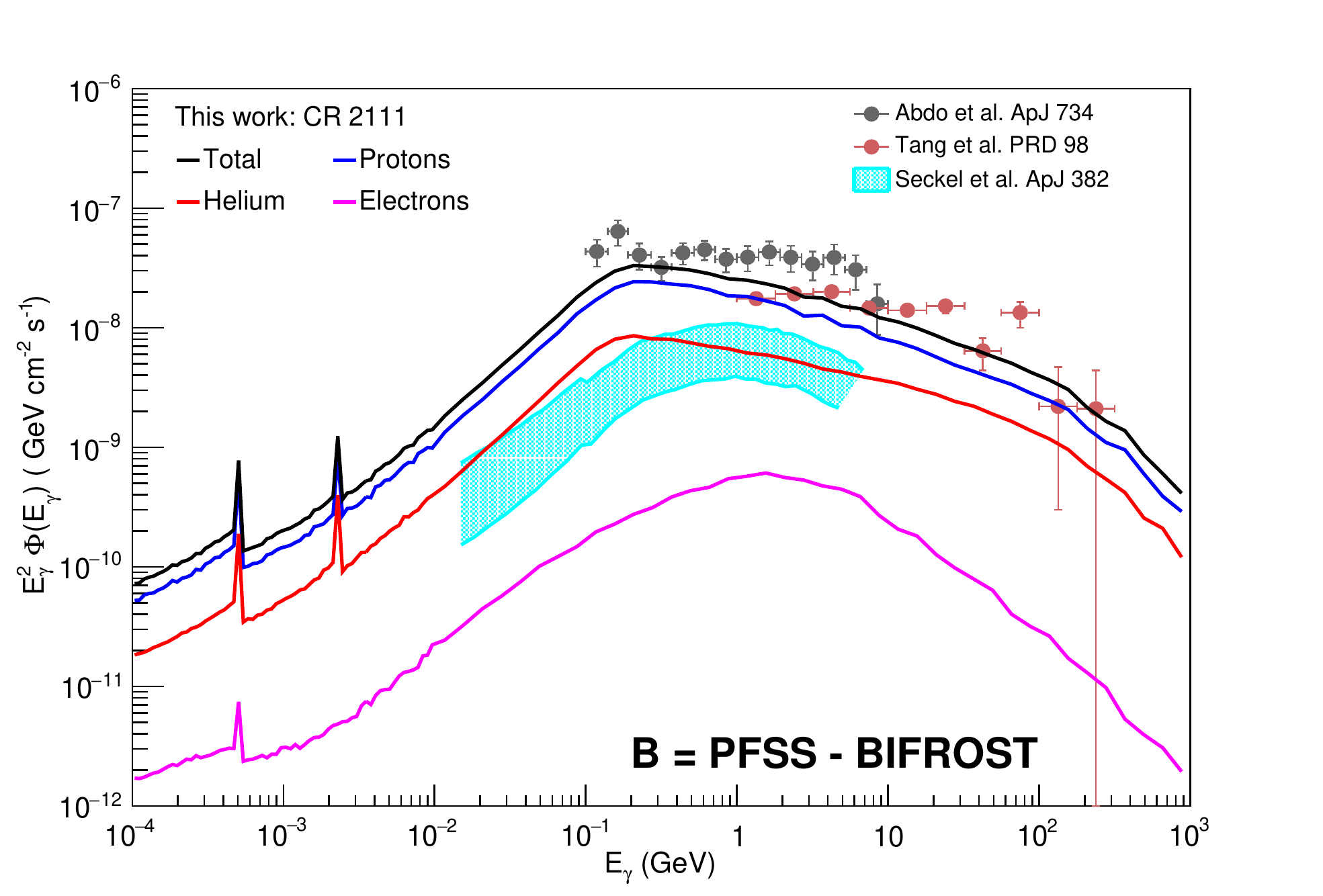}
    \caption{Gamma-ray fluxes at the Earth for the four different magnetic field configurations. Top left panel: no $B$ field; top right panel: $B$ reduced by a factor 10 with respect to the nominal PFSS configuration; bottom left panel: PFSS configuration; bottom right panel: enhanced magnetic field according to the {\tt BIFROST} model. Color lines and data points have the same meanings as n Fig.~\ref{fig:c5s}.}
    \label{fig:c5b}
\end{figure*}

\section{Effect of the magnetic field on the secondary yields}
\label{sec:beff}

The secondary emissivity of the Sun is strongly dependent on the intensity of the magnetic field close to the solar surface. To study this effect we have implemented in our simulation three additional magnetic field configurations for the CR 2111:

\begin{enumerate}
    \item $B=0$, i.e. we switch the magnetic field off;
    \item 0.1 $\times$ PFSS, i.e. we reduce the original PFSS magnetic field intensity of a factor 10;
    \item enhanced $B$ field configuration near the Sun ($r/R_\odot < 1.01$) following the {\tt BIFROST} model~\cite{2016A&A...585A...4C,Gudiksen:2011cx,hion}, i.e. we increase the original PFSS maps near the Sun to follow the {\tt BIFROST} profile~\footnote{The {\tt BIFROST} simulation is available for a limited region of the Sun. The enhancement factor is about 25 at the solar surface.}.
\end{enumerate}

Figure~\ref{fig:c5b} shows the gamma-ray fluxes at the Earth with the four different configurations of the inner magnetic field, i.e. the nominal model and the three alternative models illustrated above. 

The gamma-ray flux without magnetic field is significantly enhanced at low energies with respect to the flux in presence of magnetic field, while for gamma-ray energies above $10\units{GeV}$ the flux increases as the magnetic field increases. If the solar magnetic field is suppressed, low-energy cosmic rays can reach the Sun surface inducing a shower which produce secondary particles in the outer space. The presence of a solar magnetic field reduces the probability that low-energy cosmic rays can reach the Sun, but increases the probability of interaction for high-energy cosmic rays, since they move along curved trajectories in a strong and non-uniform magnetic field and their path length increases as the magnetic field increases. This effect is well visible when comparing the gamma-ray fluxes with $B=0$ (top left panel in Fig.~\ref{fig:c5b}) with the one with enhanced $B$ field configuration (bottom right panel in Fig.~\ref{fig:c5b})

\section{Discussion and conclusions}
\label{sec:conc}

We have implemented a full simulation with the {\tt FLUKA} code to calculate the yields of secondary particles produced by the interactions of primary cosmic rays with the solar atmosphere. Our simulation includes the current state-of-art models and data available to describe the solar atmosphere, the magnetic field nearby the Sun and in the interplanetary space.

The {\tt FLUKA} toolkit provides a detailed simulation of hadronic and electromagnetic interactions in the matter in a wide energy range, with complex geometries and even in presence of magnetic fields. The geometry used in the present work is quite flexible, and it can be used for any other configuration.

The solar atmosphere and its chemical composition have been taken from the SSM {\tt gs98} and from the model S, with some extrapolation in the chromosphere region. However, the average density should drop below $10^{-13}\units{g/cm^3}$ at an altitude of about $1400\units{km}$ from the solar surface, where the interaction probability should be negligible.  We have also used the model {\tt ags09}\footnote{\url{http://www.ice.csic.es/personal/aldos/Solar_Data_files/struct_b16_agss09.dat}} and we found very similar results.

\begin{figure}[!ht]
    \centering
    \includegraphics[width=0.95\columnwidth,height=0.23\textheight]{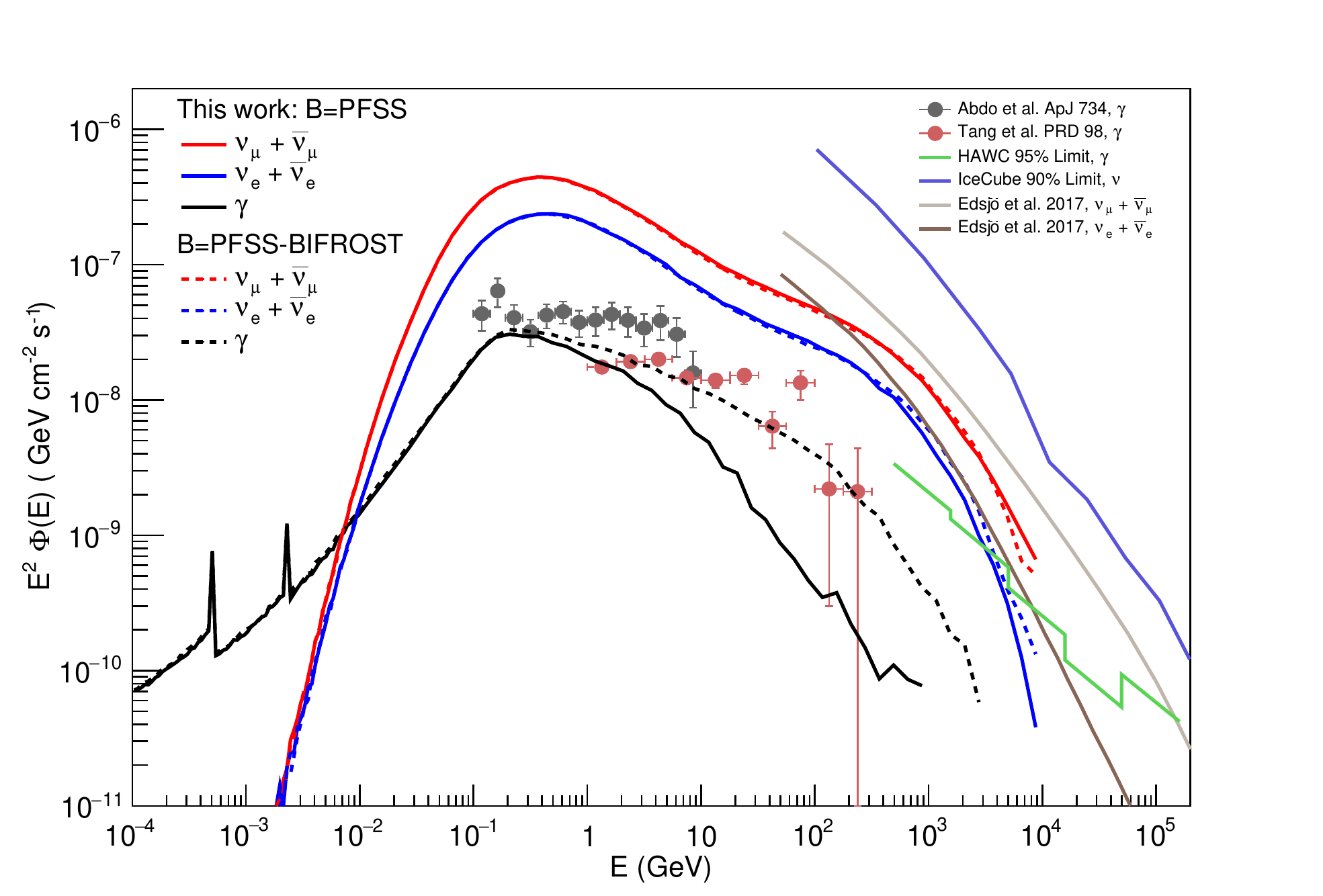}
    \caption{Gamma-ray and neutrino fluxes at the Earth calculated for the Carrington rotation 2111 with two models for the Sun magnetic field: PFSS (continuous line) and enhanced model according to the {\tt BIFROST} prediction (dashed line). Black line: gamma-ray;  red line: $\nu_\mu + \bar{\nu}_\mu$; blue line:  $\nu_e + \bar{\nu}_e$. The prediction of neutrinos by Ref.~\cite{Edsjo:2017kjk} is also shown. In addition, we include results from gamma-ray observations from Fermi-LAT~\cite{Abdo:2011xn,Tang:2018wqp}, gamma-ray HAWC’s 95\% limit~\cite{Albert:2018vcq} and neutrinos IceCube's 90\% limit~\cite{Aartsen:2019avh}.}
    \label{fig:c5_2111}
\end{figure}

The magnetic field adopted nearby the Sun is the one predicted by the PFSS model by using the synoptic map from HMI 720sline-of-sight magnetograms collected over 27-day solar rotations in a high-resolution Carrington coordinate grid. We have studied the effect of the magnetic field on the secondary yields by changing the original values of the PFSS maps. Indeed, we found that the yields are strong affected by the intensity of the magnetic field, even in the high energy region of the emission in the outer space. Recent developments of numerical solutions of a magneto-hydrodynamical (MHD) model together with the current observation of the Parker Solar Probe~\cite{Fox2016,Riley_2019} could provide new insights to get a realistic description of the plasma dynamics and of the magnetic field nearby the Sun.

The calculated solar gamma-ray flux at the Earth has been compared with the Fermi-LAT data on the disk emission in different time windows with respect to those used in the present simulations. The detected gamma-ray emission from the solar disk above 1\units{GeV} shows a harder spectrum ($\sim E_\gamma^{-2.2}$) than the cosmic-ray spectrum ($\sim E_{p,He}^{-2.7}$). This behaviour would require a high-intensity magnetic field configuration nearby the Sun, up to a factor 20 larger than the field predicted by the PFSS model used in the present simulation. 

In Fig.~\ref{fig:c5_2111} we show the predictions of our simulation about the fluxes at Earth of gamma rays and neutrinos in the CR 2111. The simulations have been performed with the standard PFSS solar magnetic field and with the enhanced one according to the {\tt BIFROST} profile. The gamma-ray production is higher in the case of the more intense magnetic field, while the effect of the magnetic field on the neutrino production seems negligible. This could be a signature that the interactions in high magnetic field occur in the higher layers of the low-density solar atmosphere, resulting in an enhanced high energy gamma-ray emission. Anyway, above 100\units{GeV} the predicted gamma-ray flux is below the HAWC's limit~\cite{Albert:2018vcq}, even in case of the enhanced solar magnetic field. The predicted neutrino flux is lower than the calculation by Ref.~\cite{Edsjo:2017kjk}, in particular below 1 \units{TeV}, due to the effect of the nearby solar magnetic field and is well below IceCube's 90\% limit~\cite{Aartsen:2019avh}. 

The magnetic field should also affect the inverse Compton gamma-ray emission close to the Sun, since the electrons should move along curved trajectories, whose lengths determine the interaction probability with the intense optical photon field. In this way the inverse Compton emission should also be peaked close to the solar surface, and might be not well separated by the disk emission due to the interaction of cosmic rays with the solar atmosphere. To get a complete picture of the solar gamma-ray emission the inverse Compton scattering needs to be calculated in presence of strong and irregular magnetic field. 
%This simulation is beyond the scope of the current paper and it deserves a dedicated work.
The current model relies on a simple calculation, in which it is assumed that electrons move along straight line~\cite{Orlando:2006zs,Orlando:2008uk,Moskalenko:2006ta}. The simulation of the inverse Compton scattering in presence of magnetic field is beyond the scope of the current paper and it deserves a dedicated work.

\begin{acknowledgments}
We acknowledge the {\tt FLUKA} collaboration for providing and supporting the code.

We acknowledge use of Joint Science Operations Center (JSOC) data.

We acknowledge use of NASA/GSFC's Space Physics Data Facility's OMNIWeb (or CDAWeb or ftp) service, and OMNI data.

We acknowledge use of simulation results provided by the Community Coordinated Modeling Center at Goddard Space Flight Center through their public Runs on Request system (\url{http://ccmc.gsfc.nasa.gov}).

We acknowledge use of simulation results provided by the  Hinode  Science  Data  Centre
Europe (\url{http://sdc.uio.no/search/simulations}).

Some of the results in this paper have been derived using the HEALPix (K.M. Górski et al., 2005, ApJ, 622, p759) package.

We thank J. Todd Hoeksema for his useful discussion on the PFSS model. 

This work has been realized using the RECAS computing infrastructure in Bari (\url{https://www.recas-bari.it/index.php/en/}). A particular acknowledgment goes to G. Donvito and A. Italiano for their valuable suport.

In this work we used custom software based on Fortran, C++, Python languages and ROOT toolkit~\cite{Brun:1997pa}.

\end{acknowledgments}

\bibliographystyle{apsrev4-1}
\bibliography{CRSun.bib}{}

\end{document}